\documentstyle{mn}
\newif\ifAMStwofonts
\input psfig.tex

\ifoldfss
  \ifCUPmtlplainloaded \else
    \NewTextAlphabet{textbfit} {cmbxti10} {}
    \NewTextAlphabet{textbfss} {cmssbx10} {}
    \NewMathAlphabet{mathbfit} {cmbxti10} {} 
    \NewMathAlphabet{mathbfss} {cmssbx10} {} 
  \fi
  \ifAMStwofonts
    \ifCUPmtlplainloaded \else
      \NewSymbolFont{upmath} {eurm10}
      \NewSymbolFont{AMSa} {msam10}
      \NewMathSymbol{\upi}     {0}{upmath}{19}
      \NewMathSymbol{\umu}     {0}{upmath}{16}
      \NewMathSymbol{\upartial}{0}{upmath}{40}
      \NewMathSymbol{\leqslant}{3}{AMSa}{36}
      \NewMathSymbol{\geqslant}{3}{AMSa}{3E}

      \let\leq=\leqslant 
       
    \fi
  \fi
\fi 

\ifnfssone
  \newmathalphabet{\mathit}
  \addtoversion{normal}{\mathit}{cmr}{m}{it}
  \addtoversion{bold}{\mathit}{cmr}{bx}{it}
  \newmathalphabet{\mathbfit} 
  \addtoversion{normal}{\mathbfit}{cmr}{bx}{it}
  \addtoversion{bold}{\mathbfit}{cmr}{bx}{it}
  \newmathalphabet{\mathbfss} 
  \addtoversion{normal}{\mathbfss}{cmss}{bx}{n}
  \addtoversion{bold}{\mathbfss}{cmss}{bx}{n}
  \ifAMStwofonts
    \ifCUPmtlplainloaded \else
      \UseAMStwoboldmath
      \makeatletter
      \new@mathgroup\upmath@group
      \define@mathgroup\mv@normal\upmath@group{eur}{m}{n}
      \define@mathgroup\mv@bold\upmath@group{eur}{b}{n}
      \edef\UPM{\hexnumber\upmath@group}
      \new@mathgroup\amsa@group
      \define@mathgroup\mv@normal\amsa@group{msa}{m}{n}
      \define@mathgroup\mv@bold\amsa@group{msa}{m}{n}
      \edef\AMSa{\hexnumber\amsa@group}
      \makeatother
      \mathchardef\upi="0\UPM19
      \mathchardef\umu="0\UPM16
      \mathchardef\upartial="0\UPM40
      \mathchardef\leqslant="3\AMSa36
      \mathchardef\geqslant="3\AMSa3E

      \let\leq=\leqslant 

    \fi
  \fi
\fi 

\ifnfsstwo
  \DeclareMathAlphabet{\mathbfit}{OT1}{cmr}{bx}{it}
  \SetMathAlphabet\mathbfit{bold}{OT1}{cmr}{bx}{it}
  \DeclareMathAlphabet{\mathbfss}{OT1}{cmss}{bx}{n}
  \SetMathAlphabet\mathbfss{bold}{OT1}{cmss}{bx}{n}
  \ifAMStwofonts
    \ifCUPmtlplainloaded \else
      \DeclareSymbolFont{UPM}{U}{eur}{m}{n}
      \SetSymbolFont{UPM}{bold}{U}{eur}{b}{n}
      \DeclareSymbolFont{AMSa}{U}{msa}{m}{n}
      \DeclareMathSymbol{\upi}{0}{UPM}{"19}
      \DeclareMathSymbol{\umu}{0}{UPM}{"16}
      \DeclareMathSymbol{\upartial}{0}{UPM}{"40}
      \DeclareMathSymbol{\leqslant}{3}{AMSa}{"36}
      \DeclareMathSymbol{\geqslant}{3}{AMSa}{"3E}

      \let\leq=\leqslant 

    \fi
  \fi
\fi 

\ifCUPmtlplainloaded \else
  \ifAMStwofonts \else 
    \def\upi{\pi}
    \def\umu{\mu}
    \def\upartial{\partial}
  \fi
\fi

\begin{document}

\title{Wide Field Imaging. I. Applications of Neural Networks \\
to object detection and star/galaxy classification}
\author[S. Andreon, G. Gargiulo, G. Longo, R. Tagliaferri and N. Capuano]
{S. Andreon$^{1}$, G. Gargiulo $^{2,3}$, G. Longo$^{1}$, R. Tagliaferri $^{4,5}$ and N. Capuano$^{2}$,\\
1. Osservatorio Astronomico di Capodimonte, via Moiariello 16, 80131 Napoli,
Italia\\
2. Facolt\`{a} di Scienze, Universit\`{a} di Salerno, via S. Allende, 84081
Baronissi (Salerno), Italia\\
3. IIASS ''E. R. Caianiello'', via G. Pellegrino, 19, 84019 Vietri s/m
(Salerno), Italia\\
4. DMI, Universit\`{a} di Salerno, via S. Allende, 84081 Baronissi
(Salerno), Italia\\
5. I.N.F.M. Unit\`{a} di Salerno, via S. Allende, 84081 Baronissi (Salerno),
Italia}
\date{Accepted .... Received .....; in original form ...... }
\maketitle

\pagerange{\pageref{firstpage}--\pageref{lastpage}}
\label{firstpage}

\begin{abstract} Astronomical Wide Field Imaging performed with new large
format CCD detectors poses data reduction problems of unprecedented scale which
are difficult to deal with  traditional interactive tools. We present here NExt
(Neural Extractor): a new Neural  Network (NN) based package capable to detect
objects and to perform both deblending and star/galaxy classification in an
automatic way. Traditionally, in astronomical images, objects are first
discriminated from the noisy background by searching for sets of connected
pixels having brightnesses above a given  threshold and then they are classified as
stars or as galaxies through diagnostic diagrams having variables choosen
accordingly to the astronomer's taste and experience. In the extraction step,
assuming that images are well sampled, NExt requires only the simplest a priori 
definition of ``what an object is"  (id est, it keeps all structures composed 
by more than one pixels) and performs the detection via an
unsupervised NN approaching detection as a clustering problem which has
been thoroughly studied in the artificial intelligence literature.   The first
part of the NExt procedure consists in an optimal compression of the redundant
information contained in the pixels via a mapping from pixels intensities to a
subspace individuated through principal component analysis. At magnitudes
fainter than the completeness limit, stars are usually  almost
indistinguishable from galaxies, and therefore the parameters characterizing
the two  classes do not lay in disconnected subspaces, thus preventing the use
of unsupervised methods. We therefore adopted a supervised NN (i.e. a NN which
first learns the rules to classify objects  from examples and then applies them
to the whole dataset).  In practice, each object is classified depending on its
membership to the regions  mapping the input feature space in the training
set.  In order to obtain an objective and reliable classification, instead of
using an arbitrarily  defined set of features,  we use a NN to select the most
significant features among the large number of measured ones,
and then we use their selected features to perform  the
classification task.  In order to optimise the performances of the system we
implemented and tested several different  models of NN.  The comparison of the
NExt performances with those of the best detection and classification package
known to the authors (SExtractor) shows that NExt is at least  as effective as
the best traditional packages.

\end{abstract}



\section{Introduction}

Astronomical wide field (hereafter WF) imaging encompasses the use of images
larger than $4000^{2}$ pxls (Lipovestky 1993) and is the only tool to tackle
problems based on rare objects or on statistically significant samples of
optically   selected objects.   Therefore, WF imaging has been and still is of
paramount  relevance to almost all field of astrophysics: from the structure and
dynamics  of our Galaxy, to the environmental effects on galaxy formation and
evolution,  to the large scale structure of the Universe.  In the past, WF was the
almost exclusive domain of Schmidt telescopes equipped  with large photographic
plates and was the main source of targets for photometric and spectroscopic
follow-up's at telescopes in the 4 meter class.  Nowadays, the exploitation of the
new generation 8 meter class telescopes, which are designed to observe targets
which are often too faint to be even detected on photographic material (the
POSS-II\ detection limit in B\ is $\sim 21.5$ $mag)$, requires digitised surveys
realized with large format CCD detectors  mounted on dedicated telescopes.  Much
effort has therefore been devoted worldwide to construct such facilities:  the
MEGACAM project at the CFH, the ESO\ Wide Field Imager at the 2.2 meter telescope,
the Sloan -\ DSS\ and the ESO-OAC\ VST (Mancini et al. 1999) being only a few
among the ongoing or planned experiments.

One aspect which is never too often stressed is the humongous problem posed by
the handling, processing and archiving of the data produced by these
instruments: the VST alone, for instance, is expected to produce a flow of
almost 30 GByte of data per night or more than 10 Tbyte per year of operation.

The scientific exploitation of such a huge amount of data calls for new data reduction
tools which must be reliable, must require a small amount of interactions with the operators
and need to be as much independent on a priori choices as possible.

In processing a WF image, the final goal is usually the construction of a
catalogue containing as many as possible astrometric, geometric, morphological
and photometric parameters for each individual object present on the image.
The first step in any catalogue construction is therefore the detection of the
objects, a step which, as soon as the quality of the images increases (both in
depth and  in resolution), becomes much less obvious than what it may seem at
first glance.  The traditional definition of ''object" as a set of connected
pixels having  brightness higher than a given threshold, has in fact several
well known pitfalls.  For instance, low surface brightness galaxies very often
escape recognition since i) their central brightness is often comparable or
fainter than the detection threshold, and  ii) their shape is clumpy, which
implies that even though there may be several  nearby pixels above the
threshold, they can often be not connected and thus escape  the assumed
definition.

A similar problem is also encountered in the catalogues extracted from the
{\it Hubble Deep Field}  (HDF) where a pletora of small ``clumpy" objects is
detected but it is not clear whether  each clump represents an individual
object or rather is a fragment of a larger one.  Ferguson (1998) stresses some
even stronger limitations of the traditional  approach to  object
detection:  i) a comparison of catalogues obtained by different groups from the
same raw material  and using the same software shows that, near the detection
limits the results are  strongly dependent on the assumed definition of
"object";  ii) object detection performed by the different groups is worse than
what even an  untrained astronomer can attain by visually inspecting an image:
many objects which are present in the image are lost by the software, while
others which are missing on the image  are detected (hereafter spurious
objects; see Fig. 1).  In other words: the silent assumption that faint objects
consist of connected and amorphous sets of pixels  makes the definition of ''object"
astronomer--dependent and produces quite ambiguous results at very low S/N
ratios.

The classification on morphological grounds only of an object as a star or as a
galaxy substantially relies on whether the object is spatially resolved or not.
Human experts can usually classify objects either directly from the appearance of
the objects on an image (either photographic or digital) or from the value of some
finite set  of derived parameters via diagnostic diagrams (such as magnitude
versus area). This approach, however, is too much time consuming and too much
dependent on  the ''know how" and personal experience of the observer: i) the
choice of the most  suitable parameters largely varies from author to author thus 
making comparisons  difficult if not at all impossible, and  ii) regardless the
complexity of the problem,  due to the obvious limitations of representing  three
or more dimensions spaces on a  two dimensional graph, only two features are often
considered.  In recent years much effort has therefore been devoted to implement
and fine tune  Artificial Intelligence (herafter AI) tools to perform star/galaxy
classification  on authomatic grounds.  The powerful package SExtractor (SEx,
Bertin \& Arnouts 1996), for instance,   uses nine features (eight isophotal areas
and the peak intensity) and a neural  network to classify objects.   The SEx
output is an index, ranging from 0 to 1, which gives the degree of  "stellarity" 
of the object.   This implies, however, still a fair degree of arbitrarity in
choosing these features and not any other set.  Other approaches to the same
problems will be reviewed in the last section of  this paper.


This paper is divided in two major parts: in \S 2 we present the AI theory used
in the paper, and in \S 3 the experiments.  Finally, in \S 4 we discuss our
results and draw the conclusions.

\section{The Theory}

In the AI domain there are dozens of different NN's used and optimised to
perform the most  various tasks. In the astronomical literature, instead, only
two types of NN's are used:  the ANN, called in the AI literature Multi-layer
Perceptron (MLP) with back-propagation learning algorithm, and the Kohonen's
Self-organizing Maps (or their supervised generalization).

We followed a rather complex approach which can be summarised as follows:
Principal Component Analysis (PCA) NN's were used to reduce the dimensionality
of the input space. Supervised NN's need a large amount of labelled data to
obtain a good classification while unsupervised NN's overcome this need, but do
not provide good performances when classes are not well separated.  Hybrid and
unsupervised hierarchical NN's are therefore very often introduced  to simplify
the expensive post-processing step of labelling the output neurons in classes
(such as objects/background), in the object detection process. In the following
subsections we illustrate the properties of several   types of NN's which were
used in one or another of the various tasks.  All the discussed models were
implemented, trained and tested and the results  of the best performing ones
are illustrated in detail in the next sections.

\subsection{PCA Neural Nets\label{section3.1}}

A pattern can be represented as a point in a $L$-dimensional parameter
space. To simplify the computations, it is often needed a more compact
description, where each pattern is described by $M$, with $M<L$,
parameters. Each $L$-dimensional vector can be written as a linear
combination of $L$ orthonormal vectors or as a smaller number of
orthonormal vectors plus a residual error. PCA is used to select the
orthonormal basis which minimizes the residual error.

Let $x$ be the $L$-dimensional zero mean input data vectors and
$C=E(xx^{T})=\left\langle xx^{T}\right\rangle $ be the covariance matrix
of the input vectors $x$. The $i$-th principal component of $x$ is defined
as $x^{T}c(i)$, where $c(i)$ is the normalized eigenvector of $C$
corresponding to the $i$-th largest eigenvalue $\lambda(i)$.

The subspace spanned by the principal eigenvectors $c(1),...,c(M),(M<L)$ is
called  the PCA subspace (with dimensionality $M$; Oja 1982; Oja et al. 1996).
In order to perform PCA, in some cases and expecially in the non linear one, it
is convenient to  use NN's which can be implemented in various ways (Baldi \&
Hornik 1989; Jutten \& Herault 1991; Oja 1982;  Oja, Ogawa \& Wangviwattana
1991; Plumbley 1993; Sanger 1989).  The PCA NN used by us was a feedforward
neural network with only one layer which is  able to extract the principal
components of the stream of input vectors.  Fig. 2 summarises the structure of
the PCA NN's. As it can be seen, there is one input layer, and one forward
layer of neurons which is totally connected to the inputs.  During the learning
phase there are feedback links among neurons, the topology of  which classifies
the network structure as either hierarchical or symmetric depending  on the
feedback connections of the output layer neurons.

Typically, Hebbian type learning rules are used, based on the one unit
learning algorithm originally proposed in (Oja 1982). The adaptation step
of the learning algorithm - in this case the network is composed by only
one output neuron - is then written as:

\begin{equation}
w_{j}^{(t+1)}=w_{j}^{(t)}+\mu y^{(t)}\left[ x_{j}^{(t)}-y^{(t)}w_{j}^{(t)}%
\right]  \label{eq2.1}
\end{equation}

where $x_{j}^{(t)}$, $w_{j}(t)$ and $y^{(t)}$ are, respectively, the value
of the $j$-th input, of the $j$-th weight and of the network output at
time $t$, while $\mu $ is the learning rate. $\mu y^{(t)}x_{j}^{(t)}$ is
the Hebbian increment and Eq.1 satisfies the condition:

\begin{equation}
\sum_{j=1}^{M}\left( w_{j}^{(t)}\right) ^{2}\leq 1  \label{eq2.2}
\end{equation}

Many different versions and extensions of this basic algorithm have been
proposed in recent years (Karhunen \& Joutsensalo 1994 = KJ94; Karhunen \&
Joutsensalo 1995 = KJ95; Oja et al. 1996; Sanger 1989).

The extension from one to more output neurons and to the hierarchical
case gives the well known Generalized Hebbian Algorithm (GHA) (Sanger 1989; KJ95):

\begin{equation}
w_{ji}^{(t+1)}=w_{ji}^{(t)}+\mu y_{j}^{(t)}\left[ x_{i}^{(t)}-%
\sum_{k=1}^{j}y_{k}^{(t)}w_{jk}^{(t)}\right]  \label{eq2.3}
\end{equation}

while the extension to the symmetric case gives the Oja's Subspace Network
(Oja 1982):

\begin{equation}
w_{ji}^{(t+1)}=w_{ji}^{(t)}+\mu y_{j}^{(t)}\left[ x_{i}^{(t)}-%
\sum_{k=1}^{M}y_{k}^{(t)}w_{jk}^{(t)}\right]  \label{eq2.4}
\end{equation}

In both cases the weight vectors must be orthonormalized and the
algorithm stops when:

\[
t_{n}=\sqrt{\sum_{j=1}^{M}\sum_{i=1}^{L}\left(
w_{ji}^{(t)}-w_{ji}^{(t-1)}\right) ^{2}} <\varepsilon
\]

where $\varepsilon $ is an arbitrarily choosen small value.  After the learning
phase, the network becomes purely feedforward.  KJ94 and KJ95 proved that PCA
neural algorithms can be derived from optimization problems,  such as variance
maximization and representation error minimization.  They  generalized these
problems to nonlinear problems, deriving nonlinear algorithms (and the relative
networks) having the same structure of the linear ones: either hierarchical or
symmetric.  These learning algorithms can be further classified in robust PCA
algorithms and nonlinear PCA algorithms.  KJ95 defined robust PCA as those in
which the objective function grows slower than a quadratic  one.  The non
linear learning function appears at selected places only.  In nonlinear PCA
algorithms all the outputs of the neurons are nonlinear function of the
responses.

More precisely, in the robust generalization of variance maximization, the
objective  function $f(z)$ is assumed to be a valid cost function such as
$ln\;cos(z)$ or $|z|$.  This leads to the adaptation step of the learning
algorithm:

\begin{equation}
w_{ji}^{(t+1)}=w_{ji}^{(t)}+\mu g\left( y_{j}^{(t)}\right) e_{ji}^{(t)}
\label{eq2.5}
\end{equation}

where:

\[
e_{ji}^{(t)}=x_{i}^{(t)}-\sum_{k=1}^{l(j)}y_{k}^{(t)}w_{jk}^{(t)}
\]

\[
g=\frac{df}{dz}
\]

In the hierarchical case $l(j)=j$. In the symmetric case $l(j)=M$, the
error vector $e_{j}^{(t)}$ becomes the same $e^{(t)}$ for all the neurons,
and Eq. \ref{eq2.5} can be compactly written as:

\begin{eqnarray}
W^{(t+1)} &=&W^{(t)}+\mu \left( I-W^{(t)}W^{(t)T}\right) xg\left(
x^{T}W^{(t)}\right)  \label{eq2.6} \\
&=&W^{(t)}+\mu e^{(t)}g\left( y^{(t)T}\right)  \nonumber
\end{eqnarray}

where $y^{(t)T}=x^{T}W^{(t)}$ is the instantaneous vector of neuron
responses at time $t$. The learning function $g$, derivative of $f$, is
applied separately to each component of the argument vector.

The robust generalisation of the representation error problem (KJ95)
with $f(t)\leq t^{2}$, leads to the stochastic gradient algorithm:

\begin{equation}
w_{j}^{(t+1)}=w_{j}^{(t)}+\mu \left( w_{j}^{(t)T}g\left( e_{j}^{(t)}\right)
x^{(t)}+x^{(t)T}w_{j}^{(t)}g\left( e_{j}^{(t)}\right) \right)  \label{eq2.7}
\end{equation}

This algorithm can be again considered in both the hierarchical and symmetric
cases. In the symmetric case $l(j)=M$, the error vector is the same $e^{(t)}$
for all the weights $w^{(t)}$. In the hierarchical case $l(j)=j$, Eq.
\ref{eq2.7} gives the robust counterparts of principal eigenvectors $c(i)$.

In Eq. \ref{eq2.7} the first update term

\[
w_{j}^{(t)T}g\left( e_{j}^{(t)}\right) x^{(t)}
\]

is proportional to the same vector $x(t)$ for all weights $w_{j}(t)$.
Furthermore, we can assume that the error vector $e(t)$ is relatively
small after the initial convergence. Hence, we can neglect the first term
in Eq. \ref{eq2.7} and this leads to:

\begin{equation}
w_{j}^{(t+1)}=w_{j}^{(t)}+\mu y_{j}^{(t)}g\left( e_{j}^{(t)}\right)
\label{eq2.8}
\end{equation}

Let us consider now the nonlinear extensions of PCA algorithms which can be
obtained in a heuristic way by requiring all neuron outputs to be always nonlinear
in Eq. \ref{eq2.5}, then:

\begin{equation}
w_{j}^{(t+1)}=w_{j}^{(t)}+\mu g\left( y_{j}^{(t)}\right) b_{j}^{(t)}
\label{eq2.9}
\end{equation}

where:

\[
b_{j}^{(t)}=x^{(t)}-\sum_{k=1}^{l(j)}g\left( y_{k}^{(t)}\right) w_{k}^{(t)}
\]

In previous experiments (Tagliaferri et al. 1999, Tagliaferri et al. 1998)
we found that the hierarchical robust NN of Eq. \ref{eq2.5} with learning
function $g=tanh(\alpha x)$ achieves the best performance with respect to
all the other mentioned PCA NN's and linear PCA.

\subsection{Unsupervised neural nets}

Unsupervised NN's partition the input space into clusters and assign to
each neuron a weight vector which univocally individuates the template
characteristic of one cluster in the input feature space. After the
learning phase, all the input patterns are classified.

Kohonen (1982, 1988) Self Organizing Maps (SOM) are composed by one neuron
layer structured in a rectangular grid of $m$ neurons. When a pattern $x$
is presented to the NN, each neuron $i$ receives the input and computes
the distance $d_{i}$ between its weight vector $w_{i}$ and $x$. The neuron
which has the minimum $d_{i}$ is the winner. The adaptation step consists
in modifying the weights of the neurons in the following way:

\begin{equation}
w_{j}^{(t+1)}=w_{j}^{(t)}+\varepsilon ^{(t)}h_{\sigma ^{(t)}}\left( d\left(
j,k\right) \right) \left( x-w_{j}^{(t)}\right)  \label{eq2.10}
\end{equation}

where $\varepsilon ^{(t)}$ is the learning rate ($0\leq \varepsilon^{(t)}\leq 1$)
decreasing in time, $d\left( j,k\right) $ is the distance in the grid between the
$j$ and the $k$ neurons and $h_{\sigma ^{(t)}}\left(x\right) $ is a unimodal
function with variance $\sigma ^{\left( t\right)
}\; $ decreasing with $x$.

The Neural-Gas NN is composed by a linear layer of neurons and a modified
learning algorithm (Martinetz Berkovitch \& Shulten 1993).
It classifies the neurons in an ordered list $\left( j_{1},...,j_{m}\right) $ accordingly to
their distance from the input pattern.
The weight adaptation depends on the position
$rank(j)$ of the $j$-th neuron in the list:

\begin{equation}
w_{j}^{(t+1)}=w_{j}^{(t)}+\varepsilon ^{(t)}h_{\sigma ^{(t)}}\left(
rank\left( j\right) \right) \left( x-w_{j}^{(t)}\right)  \label{eq2.11}
\end{equation}

and works better than the preceding one: in fact, it is quicker and
reaches a lower average distortion value\footnote{Let $P(x)$ be the
pattern probability distribution over the set $V\subseteq \Re ^{n}$ and
let $w_{i}(x)$ be the weight vector of the neuron which classifies the
pattern $x$. The average distortion is defined as $E=\int P(x)\left(
x-w_{i}(x)\right) ^{2}dx$\par {}}.

The Growing Cell Structure (GCS) (Fritzke 1994) is a NN which is capable to
change its structure depending on the data set.  Aim of the net is to map the
input pattern space into a two-dimensional discrete structure $S$ in such a way
that similar patterns are represented by topological neighboring elements. The
structure $S$ is a two-dimensional simplex where the vertices are the neurons
and the edges attain the topological information. Every modification of the net
always maintains the simplex properties. The learning algorithm starts with a
simple three node simplex and tries to obtain an optimal network by a
controlled growing process:  {\it id est}, for each pattern $x$ of the training
set, the winner $k$ and  the neighbors weights are adapted as follows:

\begin{equation}
w_{k}^{(t+1)}=w_{k}^{(t)}+\varepsilon _{b}\left( x-w_{k}^{(t)}\right)
;\;\;w_{j}^{(t+1)}=w_{j}^{(t)}+\varepsilon _{n}\left( x-w_{j}^{(t)}\right)
\label{eq2.12}
\end{equation}

$\forall j$ connected to $k$; where $\varepsilon _{b}$ and $\varepsilon _{n}$
are constants which determine the adaptation strength for the winner and for
the neighbors, respectively.

The insertion of a new node is made after a fixed number $\lambda $ of
adaptation steps. The new neuron is inserted between the most frequent
winner neuron and the more distant of its topological neighbors. The
algorithm stops when the network reaches a pre-defined number of elements.

The on-line K-means clustering algorithm (Lloyd 1982) is a simpler
algorithm which applies the Gradient Descent (=GD) directly to the average
distortion function as follows:

\begin{equation}
w_{j}^{(t+1)}=w_{j}^{(t)}+\varepsilon ^{(t)} \left( x-w_{j}^{(t)}\right)
\label{eq2.13}
\end{equation}

The main limitation of this technique is that the error function presents
many local minima which stop the learning before reaching the optimal
configuration.

Finally, the Maximum Entropy NN (Rose, Gurewitz \& Fox, 1990) applies the GD
to the error function to obtain the adaptation step:

\begin{equation}
w_{j}^{(t+1)}=w_{j}^{(t)}+\varepsilon ^{(t)} \frac{\exp \left( -\beta
^{(t)}d_{j}\right) }{\sum_{k=1}^{m}\exp \left( -\beta ^{(t)}d_{k}\right) }%
\left( x-w_{j}^{(t)}\right)  \label{eq2.14}
\end{equation}

\bigskip

where $\beta $ is the inverse temperature and takes value increasing in time
and $d_{j}$ is the distance between the $j$-th and the winner neurons.

\subsection{Hybrid neural nets}

Hybrid NN's are composed by a clustering algorithm which makes use of the
information derived by one unsupervised single layer NN. After the learning
phase of the NN, the clustering algorithm splits the output neurons in a number
of subsets which is equal to the number of the desired output classes. Since
the aim is to put similar input patterns in the same class and dissimilar input
patterns in different classes, a good strategy consists in applying a
clustering algorithm directly to the weight vectors of the unsupervised NN.

A non-neural agglomeration clustering algorithm that divides the pattern set
(in this case the weights of the neurons) $W=\left\{ w_{1},...,w_{m}\right\}
$ in $l$ clusters (with $l<m$) can be briefly summarized as follows:

\begin{enumerate}
\item[-] it initially divides $W$ in $m$ clusters $C_{1},...,C_{m}$ such
that $C_{j}=\left\{ w_{j}\right\} $;

\item[-] then it computes the distance matrix $D$ with elements $D_{ij}=d\left(
C_{i},C_{j}\right) $;

\item[-] then it finds the smallest element $D_{ij}$ and unifies the clusters
$C_{i}$ and $C_{j}$ in a new one $C_{ij}=C_{i}\cup C_{j}$;

\item[-] if the number of clusters is greater than $l$ then it goes to step 2
else, it finally stops.
\end{enumerate}

Many algorithms quoted in literature (Everitt 1977) differ only in the
way in which the distance function is computed. For example:

\[
d\left( C_{i},C_{j}\right) =\min_{w_{i{k}}\in C_{i}\;and\;w_{j{l}}\in
C_{j}}\left\| w_{i{k}}-w_{j{l}}\right\|
\]
(nearest neighbor algorithm);

\[
d\left( C_{i},C_{j}\right) =\left\| \frac{1}{\left| C_{i}\right| }%
\sum_{w_{i{k}}\in C_{i}}w_{i{k}}-\frac{1}{\left| C_{j}\right| }%
\sum_{w_{j{l}}\in C_{j}}w_{j{l}}\right\|
\]
(centroid method);

\[
d\left( C_{i},C_{j}\right) =\frac{1}{\left| C_{i}\right| \left| C_{j}\right|
}\sum_{w_{i{k}}\in C_{i},\;w_{j{l}}\in C_{j}}\left\| w_{ik}-w_{jl}\right\|
\]
(average between groups).

The output of the clustering algorithm will be a labelling of the patterns
(in this case neurons) in $l$ different classes.

\subsection{Unsupervised hierarchical neural nets}

Unsupervised hierarchical NN's add one or more unsupervised single layers NN to
any unsupervised NN, instead of a clustering algorithm as it happens in hybrid
NN's.

In this way, the second layer NN learns from the weights of the first layer
NN and clusters the neurons on the basis of a similarity measure or a distance.
The iteration of this process to a few layers gives the unsupervised hierarchical NN's.

The number of neurons at each layer decreases from the first to the output
layer and, as a consequence, the NN takes the pyramidal aspect shown in Fig.
3.  The NN takes as input a pattern $x$ and then the first layer finds the
winner neuron.  The second layer takes the first layer winner weight vector as
input and finds the second layer winner neuron and so on up to the top layer.
The activation value of the output layer neurons is 1 for the winner unit and 0
for all the others. In short: the learning steps of a $s$ layer hierarchical NN
with training set $X$ are the following:

\begin{itemize}
\item[-]  the first layer is trained on the patterns of $X$ with one of the
learning algorithms for unsupervised NN's;

\item[-]  the second layer is trained on the elements of the set $X_{2}$
which is composed by the weight vectors of the first layer winner units;

\item[-]  the process is iterated to the $i{-th}$ layer NN ($i>2$) on the
training set which is composed by the weight vectors of the winner neurons
of the $(i-1){-th}$ layer when presenting $X$ to the first layer NN,
$X_{2}$ to the second layer and so on.
\end{itemize}

By varying the learning algorithms we obtain different NN's with different
properties and abilities. For instance, by using only SOM's we have a
Multi-layer SOM (ML-SOM) (Koh J., Suk \& Bhandarkar 1995) where every layer is a two-dimensional
grid.
We can easily obtain (Tagliaferri, Capuano \& Gargiulo 1999) {\it
ML-NeuralGas}, {\it ML-Maximum Entropy} or {\it ML-K means}
organized on a hierarchy of linear layers.
The ML-GCS has a more complex
architecture and has at least 3 units for layer.

By varying the learning algorithms in the different layers, we can take
advantage from the properties of each model (for instance, since we cannot
have a ML-GCS with 2 output units we can use another NN in the output layer).


A hierarchical NN with a number of output layer neurons equal to the number
of the output classes simplifies the expensive post-processing step of
labelling the output neurons in classes, without reducing the generalization
capacity of the NN.

\subsection{Multi-layer Perceptron}

A {\it Multi-Layer Perceptron} (MLP) is a layered NN composed by:

\begin{itemize}
\item[-]  one input layer of neurons which transmit the input patterns to
the first hidden layer;

\item[-]  one or more hidden layers with units computing a nonlinear function
of their inputs;

\item[-]  and one output layer with elements calculating a linear or a
nonlinear function of their inputs.
\end{itemize}

Aim of the network is to minimize an error function which generally is the
sum of squares of the difference between the desired output (target) and the
output of the NN. The learning algorithm is called back-propagation since
the error is back-propagated in the previous layers of the NN in order to change the
weights. In formulae, let $x^{p}$ be an $L-$dimensional input vector with
corresponding target output $c^{p}$.
The error function is defined as follows:

\[
E_{p}=\frac{1}{2}\sum_{i}\left( c_{i}^{p}-y_{i}^{p}\right) ^{2}
\]

where $y_{i}^{p}$ is the output of the $i-th$ output neuron. The learning
algorithm updates the weights by using the gradient descent (GD) of the error
function with respect to the weights. If we define the input and the output
of the neuron $j$ respectively as:

\[
net_{j}^{p}=\sum_{i}w_{ji}y_{i}^{p}
\]
and
\[
y_{j}^{p}=f\left( net_{j}^{p}\right)
\]

where $w_{ji}$ is the connection weight from the neuron $i$ to the neuron $j$,
and $f\left( x\right) $ is linear or sigmoidal for the output nodes and
sigmoidal for the hidden nodes. It is well known in literature
(Bishop 1995) that these facts lead to the following adaptation steps:

\begin{equation}
w_{ji}^{(t+1)}=w_{ji}^{(t)}+\triangle w_{ji}^{(t)}\; {\rm where }\; \triangle
w_{ji}^{(t)}=\eta \delta _{j}^{p}y_{i}^{p}  \label{eq2.15}
\end{equation}

and

\begin{equation}
\delta _{j}^{p}=\left( c_{j}^{p}-y_{j}^{p}\right) y_{j}^{p}\left(
1-y_{j}^{p}\right) {\rm or } \ \delta _{j}^{p}=\left(
c_{j}^{p}-y_{j}^{p}\right) \sum_{k}\delta _{k}^{p}w_{kj}  \label{eq2.16}
\end{equation}

for the output and hidden units, respectively.
The value of the learning rate $\eta $ is small and causes a slow convergence
of the algorithm. A simple technique often used to improve it is to sum a momentum
term to Eq. \ref{eq2.15} which becomes:

\begin{equation}
\triangle w_{ji}^{(t)}=\eta \delta _{j}^{p}y_{i}^{p}+\mu \triangle
w_{ji}^{(t-1)}  \label{eq2.19}
\end{equation}

This technique generally leads to a significant improvement in the
performances of GD algorithms but it introduces a new parameter $\mu $ which has to be
empirically chosen and tuned.

Bishop (1995) and Press  et al. (1993) summarize several methods to overcome
the problems related to the local minima and to the slow time
convergence of the above algorithm.
In a preliminary step of our experiments, we tried all
the algorithms discussed in chapter 7 of Bishop (1995) finding that a
hybrid algorithm based on the scaled conjugate gradient for the first
steps and on the Newton method for the next ones, gives the best results
with respect to both computing time and relative number of errors.
In this paper we used it in the MLP's experiments.

\section{The Experiments}

\subsection{The data}

In this work we use a 2000x2000 arcsec$^{2}$ area centered on the North Galactic
Pole extracted from the slightly compressed POSS-II F plate n. 443, available via
network at the Canadian Astronomy Data Center (http://cadcwww.dao.nrc.ca). POSS-II
data were linearised using the sensitometric spots recorded on the plate. The
seeing FWHM\ of our data was 3 arcsec. The same area has been widely studied by
others and, in particular, by Infante \& Pritchet (1992, =IP92) and Infante,
Pritchet \& Hertling (1995) who used deep observations obtained at the 3.6 m CFHT
telescope in the $F$ photographic band under good seeing conditions (FWHM $<1$
arcsec) to derive a catalogue of objects complete down to $m_{F}\sim 23$, id est,
much deeper than the completeness limit of our plate. Their catalogue is therefore
based on data of much better quality and accuracy than ours, and it was for the
availability of such good template that we decided to use this region for our
experiments. We also studied a second region in the Coma cluster (which happens to
be in the same n. 443 plate) but since none of the catalogues available in
literature is much better than our data, we were forced to neglect it in most of
the following discussion.

The characteristics of the selected region, a relatively empty one, slightly
penalise our NN detection algorithms which can easily recognise objects of quite
different sizes. On the contrary of what happens to other algorithms NExt works
well even on areas where both very large and very small objects are present such
as, for instance, the centers of nearby clusters of galaxies as our preliminary
test on a portion of the Coma clusters clearly shows (Tagliaferri et al. 1998).

\subsection{Structure of NExt}
The detection and classification of the objects are a multi-step task:

1) First of all, following a widely used AI approach, we mathematically transform
the detection task into a classification one by compressing the redundant
information contained in nearby pixels by means of a non--linear PCA NN's.
Principal vectors of the PCA are computed by the NN on a portion of the whole
image. The values of the pixels in the transformed $M$ dimensional eigen-space
obtained via the principal vectors of the PCA NN are then used as inputs to
unsupervised NN's to classify pixels in few classes. We wish to stress that, in
this step, we are still classifying pixels, and not objects.


The adopted NN is unsupervised, i.e. we never feed into the detection algorithm
any a priori definition of what an object is, and we leave it free to find its own
object definition. It turns out that image pixels are split in few classes, one
coincident with what astronomers call background and some others for the objects
(in the astronomical sense). Afterwords, the class containing the background
pixels is kept separated from the other classes which are instead merged together.
Therefore, as final output, the pixels in the image are divided in ``object" or
``background".

2) Since objects are seldom isolated in the sky, we need a method to
recognise overlapping objects and deblend them. We
adopt a generalisation of the method used by Focas (Jarvis \& Tyson 1981).

3) Due to the noise, object edges are quite irregular.
We therefore apply a contour regularisation to the edges of the objects
in order to improve the following star/galaxy classification step.

4) We define and measure the features used, or suitable, for the star/galaxy
classification, then we choose the best performing features for the classification
step, through the sequential backward elimination strategy (Bishop 1995).

5) We then use a subset of the IP92 catalog to learn, validate and test the
classification performed by NExt on our images. The training
set was used to train the NN, while the validation was used for model
selection, i.e. to select the most performing parameters using an
independent data set. As template classifier, we used SEx, whose
classifier is also based on NNs.

The detection and classification performances of our algorithm were then compared
with those of traditional algorithms, such as SEx. We wish to stress that in both
the detection and classification phases, we were not interested in knowing how
well NExt can reproduce SEx or the astronomer's eye performances, but rather to
see whether the SEx and NExt catalogs are or are not similar to the ``true",
represented in our case by the IP92 catalog.

Finally, we would like to stress that in statistical pattern recognition, one of
the main problems in evaluating the system performances is the optimisation of
all the compared systems in order not to give any unfair advantage to one of the
systems with respect to the others (just because it is better optimised than the
others). For instance, since the background subtraction is crucial to the
detection, all algorithms, including SEx, were run on the same background
subtracted image.

\subsection{Segmentation}

From the astronomical point of view, segmentation allows to disentangle objects
from noisy background. From a mathematical point of view, instead, the segmentation
of an image $F$ consists in splitting it into disconnected homogeneous
(accordingly to a uniformity predicate $P$) regions $\left\{
S_{1,}...,S_{n}\right\} $, in such a way that their union is not homogeneous:

\[
\bigcup_{i=1}^{n}S_{i}=F\;\; {\rm with}\;\;S_{i}\cap S_{j}=\emptyset
,\;\;i\neq j
\]

where $P(S_{i})=true$ $\forall i$ and $P(S_{i}\cup S_{j})=false$ when
$S_{i}$ is adjacent to $S_{j}$. The two regions are adjacent when they
share a boundary, i.e. when they are neighbours.

A segmentation problem can be easily transformed into a classification one
if classes are defined on pixels and $P$ is written in such a way that
$P(S_{i})=true$ if and only if all the pixels of $S_{i}$ belong to the same
class.
For instance, the segmentation of an astronomical image in background
and objects leads to assign each pixel to one of the two classes.
Among the various methods discussed in the literature,
unsupervised NN's usually provide better performance than any other NN type on noisy
data (Pal \& Pal 1993) and have the great advantage of not requiring a definition
(or exhaustive examples) of object.

The first step of the segmentation process consists in creating a numerical
mask where different values discriminate the background from the object (Fig. 5).

In well sampled images, the attribution of a pixel to either the background or to
the object classes depends on both the pixel value and on the properties
of its neighbours: for instance, a ``bright'' isolated pixel in a
``dark'' environment is usually just noise.
Therefore, in order to classify a pixel, we need to take into account the
properties of all the pixels in a $( n\times n )$ window centered on it.
This approach can be easily extended to the case of multiband images.
$n\times n$, however, is a too high dimensionality to be effectively handled
(in terms of learning and computing time) by any classification
algorithm. Therefore, in order to lower the dimensionality, we first use a PCA to
identify the $M$ (with $M << n\times n$) most significant features.
In detail:

i) we first run the $\left( n\times n\right)$ window on a sub-image containing
representative parts of the image. We used both a $3\times3$ and a $5\times5$
windows.

ii) Then we train the PCA NN's on these patterns.
The result is a projection
matrix $W$ with dimensionality $\left( n\times n\right) \times M$, which
allows us to reduce the input feature number from $\left( n\times
n\right) $ to $M$. We considered only the first three components since, accordingly to the
PCA, they contain almost $93\%$\ of the information while the remaining $7\%$\ is distributed
over all the others.

iii) The $M$-dimensional projected vector $W\bullet I$ is the input of a
second NN which classifies the pixels in the various classes.

iv) Finally, we merge all classes except the background one in order
to reduce the classification problem to the usual ''object/background'' dichotomy.


Much attention has also to be paid to the choice of the type of PCA.
After several experiments, we found that - for our specific task which is
characterised by a large dynamical range in the luminosities of the
objects (or, which is the same, in the pixel values) -
PCA's can be split into two gross groups: PCA's with linear
input-output mapping (hereafter linear PCA NN's) and PCA's with non linear
input-output mapping (non-linear PCA NN's) (see section \ref{section3.1}).
Linear PCA NN's turned out to misclassify faint objects as background.
Non-linear PCA NN's based on a sigmoidal function allowed, instead, the
detection of faint sources.
This can be better understood from Fig. 6 and 7 which give the
distributions of the training points in the simpler case of two
dimensional inputs for the two types of PCA NN's.




Linear PCA NN's produce distributions with a very dense core (background
and faint objects) and only a few points (luminous objects) spread over a
wide area. Such a behaviour results from the presence of very luminous
objects in the training set which compress the faint ones to the bottom of
the scale.
This problem can be circumvented by avoiding very luminous objects in the training
set, but this would make the whole procedure too much dependent on the choice of the training set.

Non-linear PCA NN's, instead, produce better sampled distributions and a
better contrast between background and faint objects. The sigmoidal
function compresses the dynamical range squeezing the very luminous
objects into a narrow region (see Fig. 7).


Among all, the best performing NN (Tagliaferri et al. 1998) turned out to be the hierarchical robust
PCA NN with learning function $g^{(t)}=tanh(\alpha x)$ given in Eq. \ref{eq2.5}.
This NN was also the faster among the other non-linear PCA\ NN's.

The principal components matrices are detailed in the tables 1-3 and 4-6
for the $3 \times 3$ and $5 \times 5$ cases, respectively.
In tables 1--3, numbers are rounded to the closest integere since they differ
from an integer only at the 7-th decimal figure.
Not surprisingly, the first component turns out to be the mean in the $3
\times 3$ case. The other two matrices can be seen as anti-symmetric
filters with respect to the centre. The convolution of these filters (see
Fig. 8) with the input image gives images where the objects are the
regions of high contrast. Similar results are obtained for the $5 \times 5$
case.

At this stage we have the principal vectors and, for each pixel, we can
compute the values of the projection of each pixel in the eigenvector
space. The second step of the segmentation process consists in using
unsupervised NN's to classify the pixels into few classes, having as input
the reduced input patterns which have been just computed. Supervised NN
would require a training set specifying, for each pixel, whether that
pixel belongs to an object or to the background. We no longer consider
such a possibility, due to the arbitrariness of such a choice at low
fluxes, the lack of elegance of the method and the problems which are encountered in
the labelling phase. Unsupervised NN's are therefore necessary. We
considered several types of NN's.

As already mentioned several times,
our final goal is to classify the image pixels in just two classes:
objects and background, which should correspond to two output neurons.
This simple model, however, seldom suffice to reproduce real data in the bidimensional
case (but similar results are obtained also for the 3-D or multi-D cases),
since any unsupervised algorithm fails to produce spatially well separated clusters and more
classes are needed. A trial and error procedure shows that a good choice
of classes is $6$: fewer classes produce poor classifications while more
classes produce noisy ones. In all cases, only one class (containing the
lowest luminosity pixels) represents the background, while the other
classes represent different regions in the objects images.

We compared hierarchical, hybrid and unsupervised NN's with $6$ output neurons.
From theoretical considerations and from preliminary work (Tagliaferri et. al
1998) we decided to consider only the best performing NN's, id est Neural gas,
ML-Neural gas, ML-SOM, and GCS+ML-Neural gas. For a more quantitative and detailed
discussion see section \ref{section3.5}, where the performances of these NN's are
evaluated.

After this stage all pixels are classified in one of six classes. We merge
together all classes, with the exception of the background one and reduce
the classification to the usual astronomical dichotomy: object or background.

Finally, we create the masks, each one identifying one structure composed
by one or more objects. This task is accomplished by a simple algorithm,
which, while scanning the image row by row, when it finds one or more adjacent
pixels belonging to the object class expands the structure including all
equally labelled pixels adjacent to them.

Once objects have been identified we measure a first set of parameters. Namely:
the photometric barycenter of the objects computed as:

\[
\bar{x}=\frac{\sum_{\left( x,y\right) \in A}x\cdot I\left( x,y\right) }{%
flux }\; {\rm and \ }\bar{y}=%
\frac{\sum_{\left( x,y\right) \in A}y\cdot I\left( x,y\right) }{ flux}
\]

where $A$ is the set of pixels assigned to the object in the mask,
$I\left(x,y\right) $ is the intensity of the pixel $\left( x,y\right)$, and

\[
flux = \sum_{\left( x,y\right) \in A}I\left( x,y\right)
\]
is the flux of the object integrated over the considered area.
The semimajor axis of the object contour defined as:\

\[
a=\max_{\left( x,y\right) \in A}\left\| (x,y)-(\bar{x},\bar{y})\right\|
=\max_{\left( x,y\right) \in A}r\left( x,y\right)
\]


with position angle defined as:\
\[
\alpha =\arctan \left( \frac{y^{\prime }-\bar{y}}{x^{\prime }-\bar{x}}%
\right)
\]

where $\left( x^{\prime },y^{\prime }\right) $ is the most distant pixel
from the barycenter belonging to the object.
The semiminor axis of the faintest isophote is given by:\

\[
b=\max_{\left( x,y\right) \in A}\left| \sin \left[ \arctan \left( \frac{y-%
\bar{y}}{x-\bar{x}}\right) -\alpha \right] \right| \cdot r(x,y)
\]

These parameters are needed in order to disentangle overlapping objects.

\subsection{Object deblending}

Our method recognises multiple objects by the presence of multiple peaks in the light
distribution. Search for double peaks is performed along directions at
position angles $\beta _{i}=\alpha +i\pi /n$ with $0\leq i\leq n$.
At difference with FOCAS, (Jarvis \& Tyson 1981), we sample several position
angles because not always objects are aligned along the major axis of
their light distribution, as FOCAS implicitly assumes.
In our experiments the maximum $n$ was set to $5$.
When a double peak is found, the object is
split into two components by cutting it perpendicularly to the line
joining the two peaks.

Spurious peaks can also be produced by noise fluctuations, a case which is
very common in photographic plates near saturated objects. A good way to
minimise such noise effects is, just for deblending purposes, to reduce
the dynamical range of the pixels values, by rounding the intensity
(or pixel values) in $N$ equi-espaced levels.

Multiple (i.e. $3$ or more components) objects pose a more complex
problem. In the case shown in Fig. 9, the segmentation mask includes three
partially overlapping sources.
The search for double peaks produces a
first split of the mask into two components which separate the third and
faintest component into two fragments.
Subsequent iterations would usually produce a set of four independent components
therefore introducing a spurious detection.
In order to solve the problem posed by multiple "non spurious'' objects
erroneously split, a recomposition loop needs to be run.
Most celestial objects - does not matter whether resolved or unresolved - present a light
distribution rapidly decreasing outwards from the centre.
If an object has been erroneously split into several components, then the adjacent
pixels on the corresponding sides of the two masks will have very
different values.
The implemented algorithm checks each component (starting
from the one with the highest average luminosity and proceeding to the
fainter ones) against the others. Let us now consider two parts of an
erroneously split object. When the edge pixels have luminosity higher than
the average luminosity of the faintest component, the two parts are
recomposed. This procedure also takes care of all spurious components
produced by the haloes of bright objects (an artifact which is a major
shortcoming of many packages available in the astronomical community).

\subsection{Contour regularisation}

The last operation before measuring the objects parameters
consists in the regularization of the contours since -- due to noise, overlapping
images, image defects, etc. -- segmentation produces patterns that are not similar
to the original celestial objects that they must represent.
For the contour regularisation, we threshold the image at
several sigma over the background and we then expand the ellipse
describing the objects in order to include the whole area measured in the
object detection.


\subsection{Results on the object detection phase\label{section3.5}}

After the above described steps, it becomes possible to measure and
compare the performances of the various NN models.
We implemented and compared: Neural Gas (NG3), ML-Neural
Gas (MLNG3 or MLNG5), ML-SOM (K5), GCS+ML-Neural Gas (NGCS5).
The last digit in the NN name indicating the dimensions of the running window.

Attention was paid in choosing the training set, which needed to be at the
same time small but significant.
By trial and error, we found that for PCA NN's and unsupervised NN's
it was enough to choose $\sim 10$ sub-images,
each one $\sim 50 \times 50$ pixels wide and not containing very large
objects.
As all the experienced users know, the choice of the SEx parameters
(minimum area, threshold in units of the background noise, and deblending parameter)
is not critical and the default values were choosen (4 pixel area, $1.5\sigma$).

Table 7 shows the number of objects detected by the five NN's and SEx.
It has to be stressed that $\sim 2100$ objects out of the 4819 available in
the IP92 reference catalogue are beyond the detection limit of our
plate material.
SEx detects a larger number of objects but many of them (see Table 7)
are spurious. NN's detect a slightly smaller number of objects but most of them are
real.
In particular: MNG5 looses, with respect to SEx, only 79 real objects but
detects 400 spurious objects less; MNG3 is a little less performing in
detecting true objects but is even cleaner of spurious detections.

The upper panel of Fig. 10 shows the number of "True'' objects (i.e. objects in the
IP92 catalogue).
Most of them are fainter than $m_F \sim 21.5$ mag, id est they are fainter than the
plate limit.
The lower panel shows instead the number of objects detected by the various NN's
relative to SEx. The curves largely coincide and, in particular, MLNG5 and SEx do not statistically differ in
any magnitude bin while MLNG3 slightly differs only in the faintest bin
($m_F \sim 21.5$).

The class of ``Missed'' objects (id est objects which are listed in the reference
catalogue but are not in the NN's or SEx catalogues) needs a detailed discussion.
We focus first on brighter objects. They can be divided in:

-- Few ``True'' objects with a nearby companion which are blended in our image but
are resolved in IP92.

-- Parts of isolated single large objects incorrectly split by
IP92. A few cases.

-- A few detections aligned in the E-W direction on the two sides of the
images of a bright star. They are likely false objects (diffraction spikes
detected as individual objects in the IP92 catalog).

-- Objects in IP92 which correspond to empty regions in our images:
they can be missing because variable, fast moving, or with an overestimated
luminosity in the reference catalog; they can also be missed because spurious
in the template catalog.

Therefore, a fair fraction of the ``Missed'' objects is truly non
existent and the performances of our detection tools are lower bounded at
$m_{F}<21$ mag.
We wish to stress here that even though there is nothing like a perfect
catalogue, the IP92 template is among the best ones ever produced to our
knowledge.

The upper panel of Fig. 11 is the same as in Fig. 10. The lower panel shows
instead the fraction of "false'' objects, id est of the objects detected by the
algorithms but not present in the reference catalogue. IP92 were interested to
faint objects and masked out the bright ones, therefore their catalogue may
exclude a few ``True'' objects (in particular at $m_F \sim 17$). We believe that
all objects brighter than $m_{F}=20$ mag are really ``True'' since they are
detected both by SEx and NN's with high significance. For objects brighter than
$m_{F}=20$ mag, the NN's and SEx have similar performances. They differ only at
fainter magnitudes. The catalogue with the largest contamination by ``False''
objects is SEx, followed by MLNG5, MLNG3 and the other NN's beeing much less
contaminated. MLNG5 is quite efficient in detecting ``True'' objects and has a
$20\%\;$ cleaner detection rate in the highly populous bin $m_{F}=21.7$ mag. MLNG3
is less efficient than MLNG5 in detecting ``True'' objects but it is even cleaner
than MLNG5 of false detections.

Let us now consider whether or not the detection efficiency depends on the
degree of concentration of the light (stars have steeper central gradients than
galaxies). In IP92 objects are classified in two
major classes, star \& galaxies, and a few minor ones (merged, noise,
spike, defects, etc.) that we neglect. The efficiency of the detection is
shown in Fig. 12 for three representative detection algorithms: MLNG5, K5,
and SEx. At $m_F < 21$ mag, the detection efficiency is large, close to 1
and independent on the central concentration of the light. Please note
that there are no objects in the image having $m_F <16$ mag and that in
the first bin there are only 4 galaxies. At fainter magnitudes ($\sim
22-23$ mag) detection efficiencies differ as a function of both the
algorithm and of the light concentration. In fact, SEx, MLNG5, and to less
extent K5, turn out to be more efficient in detecting galaxies rather than
stars (in other words: ``Missed'' objects are preferentially stars). For
SEx, a possible explanation is that a minimal area above the background is
required in order for the object to be detected and at $m_F \sim 22-23$
mag and noise fluctuations can affect the isophotal area of unresolved
objects bringing it below the assumed threshold (4 pixels). This bias is
minimum for the K5 NN. However, this is more likely due to the fact that
K5 misses more galaxies than the other algorithms, rather than to the fact
that it detects more stars.


In conclusion: MLNG3 and MLNG5 turn out to have high performances 
in detecting objects: they produce catalogs which are cleaner of false
detections at the price of a slightly larger uncompleteness than
the SEx catalogues below the plate completness magnitude.

We also want to stress that since the less performing NN's produce
catalogs which are much cleaner of false detections, the selected objects
are in large part ''true'', and not just noise fluctuations.
These NN's can therefore be very suitable to select candidates for possible follow--up
detailed studies at magnitudes where many of the objects detected by SEx
would be spurious.
Deeper catalogs having a large number of spurious source, such as those
produced by SEx or other packages are instead preferable if, for instance,
they can be cleaned by subsequent processing (for instance by matching the
detected objects with other catalogs).

A posteriori, one could argue that performances similar to those of each
of the NN's could be achieved by running SEx with appropriate settings.
However, it would be unfair (and methodologically wrong) to make a fine
tuning of any of the detection algorithms using an a-posteriori knowledge.
It would also make cumbersome the authomatic processing of the images
which is the final goal of our procedure.

\subsection{Feature extraction and selection}

In this section we discuss the feature extraction and selection of the
features which are useful for the star/galaxy classification. Features are
chosen from the literature (Jarvis \& Tyson 1981; Miller \& Coe 1996; 
Odewahn et al. 1992 (=O92),
Godwin \& Peach 1977), and then selected by a sequential forward selection
process (Bishop 1995), in order to extract the most performing ones for
classification purposes.

The first five features are those defined in the previous section and
describing the ellipses circumscribing the objects: the photometric
barycenter coordinates ($\bar{x},\bar{y}$), the semimajor axis ($a$), the
semiminor axis ($b$). and the position angle ($\alpha $).
The sixth one is the object area, $A$, i.e. the number of pixels forming the object.

The next twelve features have been inspired to the pioneeristic work by
O92:
the object diameter ($dia=2a$),
the ellipticity ($ell=1-b/a$),
the average surface brightness ($\left\langle SuBr\right\rangle =\frac{1}{A}\sum_{\left(
x,y\right) \in A}I\left( x,y\right) $),
the central intensity
($I_{0}=I\left( \bar{x},\bar{y}\right) $),
the filling factor ($f_{fac}=\pi ab/A$),
the area logarithm ($c_{2}=\log \left( A\right) $),
the harmonic radius ($r_{-1}$). The latter beeing defined as:
\[
r_{-1}=\frac{1}{ flux }%
\sum_{\left( x,y\right) \in A}\frac{I\left( x,y\right) }{r\left( x,y\right) }
\]

and five gradients $G_{14}$, $G_{13}$, $G_{12}$, $G_{23}$ and $G_{34}$
defined as:
\[
G_{ij}=\frac{T_{j}-T_{i}}{r_{i}-r_{j}}
\]

where $T_{i}$ is the average surface brightness within an ellipse, with
position angle $\alpha $, semimajor axis $r_{i}=i\,a/4$, $i=1,...4$.
and ellipticity $ell$.



Two more features are added following Miller \& Coe (1996): the ratios
$T_{r}=\left\langle SuBr\right\rangle /I_{0}$ and $T_{cA}=I_{0}/\sqrt{A}$.

Finally, five FOCAS features (Jarvis \& Tyson 1981) have been included:
the second ($C_{2}$) and the fourth ($C_{4}$) total moments defined as:
\[
C_{2}=\frac{M_{20}+M_{02}}{M_{00}}\qquad {\rm and}\qquad C_{4}=\frac{%
M_{40}+2M_{22}+M_{04}}{M_{00}}
\]

where $M_{ij}$ are the object central momenta computed as:
\[
M_{ij}=\sum_{\left( x,y\right) \in A}\left( x-\bar{x}\right) ^{i}\left( y-%
\bar{y}\right) ^{j}I\left( x,y\right) ,
\]

the average ellipticity:
\[
E=\frac{\sqrt{\left( M_{20}-M_{02}\right) ^{2}+M_{11}^{2}}}{M_{02}+M_{20}},
\]

the central intensity averaged in a $3\times 3$ $area$  and, finally, the
''Kron'' radius defined as:
\[
r_{Kron}=\frac{1}{ flux }%
\sum_{\left( x,y\right) \in A}I\left( x,y\right) r\left( x,y\right)
\]

For each object we therefore measure $25$ features, where the first $6$ are
reported only to easy the graphical representation of the objects and have a low
discriminating power. The complete set of the extracted features is given in Table
8.

Our list of features includes therefore most of those usually used in the
astronomical literature for the star/galaxy classification.

Are all these features truly needed? And, if this is not and a smaller
subset contains all the needed information, what are the most useful ones?
We tried to answer these questions by evaluating the classification
performance of each set of features through the a-priori knowledge of the
true classification of each object, as it is listed in a much deeper and
higher quality reference catalog.

Most of the defined features are not independent. The presence of redundant
features decreases the classification performances since any algorithm would try
to minimise the error with respect to features which are not particularly relevant
for the task. Furthermore, by introducing useless features the computational speed
would be lowered.

The feature selection phase was realised through the sequential backward
elimination strategy (Bishop 1995), which works as follows: let us suppose to have
$M$ features in one set and to run the classification phase with this set. Then,
we build $M$ different sets with $M-1$ features in each one and then we run the
classification phase for each set, keeping the set which attains the best
classification. This procedure allows us to eliminate the less significant
feature. Then, we repeat $M-1$ times the procedure eliminating one feature at each
step. In order to further reduce the computation time we do not use the validation
set and the classification error is evaluated directly on the test set. It has to
be stressed that this procedure is common in the statistical pattern recognition
literature where, very often, for this task are also introduced simplified models.
This however could be avoided in our case due to the speed and good performances
of our NN's

Unsupervised NN's were not successful in this task, because the input data
feature space is not separated into two not overlapping classes (or, in
simpler terms, the images and therefore the parameters of stars and
galaxies fainter than the completeness limit of the image are quite
similar), and they reach a performance much lower than supervised NN's.

Supervised learning NN's give far better results. We used a MLP with one
hidden layer of $19$ neurons and only one output, assuming value $0$ for
star and value $1$ for galaxy. After the training, we calculate the NN
output as $1$ if it is greater than $0.5$ and $0$ otherwise for each
pattern of the test set. The experiments produce a series of catalogues,
one for each set of features.

Fig. 13 shows the classification performances as a function of the adopted
features. After the first step, the classification error remains almost constant
up to $n=4$, id est up to the point where features which are important for the
classification are removed.

A high performance can be reached using just 6 features. With a lower number of 
features the classification worsen, whereas a larger number of features is
unjustified, because it does not increase the performances of the system. The best
performing set of features consists of features 11, 12, 14, 19, 21, 25 of table 8.
They are two radii, two gradients, the second total moment and a ratio which
involves measures of intensity and area.



%
%

\subsection{Star/Galaxy classification}

Let us discuss now how the star/galaxy classification takes place. The first step
is accomplished by ``teaching" the MLP NN using the selected best features. In
this case we divided the data set into three independent data sets: training,
validation and test sets.  The learning optimization is performed using the
training set while the early stopping technique (Bishop 1995) is used on the
validation set to stop the learning to avoid overfitting. Finally, we run the MLP
NN on the test set.

As comparison classifier, we adopt SEx, which is based on a MLP NN. As
features useful for the classification, SEx uses eight isophotal areas and
the peak intensity plus a parameter, the FWHM of stars. Since the SEx NN
training was already realised by Bertin \& Arnouts (1996) on $10^6$ simulated
images of stars and galaxies, we limit ourselves to tune SEx in order to
obtain the best performances on the validation set. Both SEx and our system
use NN's for the classification, but they follow two different, alternative
approaches: SEx uses a very large training set of simulated stars and
galaxies, our system uses noisy, real data. Furthermore, while the features
of SEx are fixed by the authors, and the NN's output is a number $x$,
$0<x<1$;  our system selects the best performing ones and its output is an
integer: 0 or 1 (id est star or galaxy). Therefore, we use the validation set
for choosing the threshold which maximises the number of correct
classifications by SEx (see Fig. 14).

The experimental results are shown in Fig. 15 where the errors are plotted as
a function of the magnitude. At all magnitudes NExt misclassify less objects
than SEx. Out of 460 objects, SEx makes 41 incorrect classifications, while
NExt just 28. 

In order to check the that our feature selection is optimal, we also compared
our classification with those obtained using our MLP NN's with others feature
sets, selected as shown in Table 9. The total number of misclassified objects
in the test set of 460 elements were: O-F, 43 errors; O-L, 30 errors; O-S, 35
errors; GP1, 48 errors; GP2, 49 errors.
Fig. 16 shows the classification performances of the considered feature
sets as a function of the magnitude of the objects.
Results for stars are presented as solid line, while for galaxies we used
dotted lines. The perfomances of NExt are presented in the top-left panel:
galaxies are correctly classified as long as they are detected, whereas the
correctness of the classification of stars drops to 0 at $m_F=21$. Fainter
stars are pratically absent in the IP92 catalog, thus explaining why the
stars point stop at brighter magnitudes than galaxies. O92
selected a 9 features set (O-F) for the star/galaxy classification.
Their set (central--left panel) is slightly less performing for bright
($m_F=17$) galaxies and for faint stars ($m_F=19$) than the set of features
selected by us (upper--left panel). They select also a smaller (four) set of
features (O-F) quite useful to classify large objects. The classification
performances of this set, when applied to our images, turn out to be better
than the larger feature dataset: in fact, bright galaxies are not
misclassified (see the bottom left panel). Even with respect to our dataset
O-F performs well: their set is sligthly better in classifying bright
galaxies, at the price of a achieving lower performances on faint stars. The
further set of features by O92 (O-S) was aimed to the accurate detection of
faint sources and performs similarly to their full set: it misclassifies
bright galaxies and faint stars. The performances of the traditional
classifiers, $mag \ vs \ area$ (GP1) and $mag \ vs \ brightness$ (GP2), are
presented in the central and low right panels. With just two features, all
the faint objects are classified as galaxies, and due to the absence of stars
in our reference catalog, the classification performances are $100 \%$.
However, this is not a real classification. At bright magnitudes, the
classification of the traditional classified dataset are as large as, or
sligthly lower, than the NExt dataset.

\section{Summary and conclusions}


In this paper we discuss a novel approach to the problem of detection  and
classification of objects on WF images. In Section 2 we shortly review the theory
of  some type of NN's which are not familiar to  the astronomical community. Based
on these considerations, we implemented a Neural Network  based procedure (NExt)
capable to perform the following tasks:  i) to detect objects against a noisy
background; ii) to deblend  partially overlapping objects; iii) to separate stars
from galaxies.  This is achieved by a combination of three different NN's each
performing a specific task. First we run a non linear PCA NN to reduce the
dimensionality of the input  space via a mapping from pixels intensities to a
subspace individuated through principal component analysis. For the second step we
implemented a hierarchical unsupervised NN to segmentate the image and, finally
after a deblending and reconstruction loop we implemented a supervised MLP to 
separate stars from galaxies. 

In order to identify the best
performing NN's we implemented and tested in  homogeneous conditions several
different models.  NExt offers several methodological and practical advantages
with respect to other  packages: i) it requires only the simplest a priori
definition of what an "object" is; ii) it uses unsupervised algorithms for all
those tasks where both theory and  extensive testing show that there is no loss in
accuracy with respect to  supervised methods. Supervised methods are in fact used
only to  perform star/galaxy separation since, at magnitudes fainter than the
completeness  limit, stars are usually  almost indistinguishable from galaxies and
the parameters  characterizing the two  classes do not lay in disconnected
subspaces. iii) Instead of using an arbitrarily defined and often specifically
tailored set of features  for the classification task NExt, after measuring a
large set of geometric and photometric  parameters, uses a sequential backward
elimination strategy (Bishop 1995) to select only the  most significant ones. The
optimal selection of the features was checked against  the performances of other
classificators (see Sect. 3.8).
 
In order to evaluate the performances of NExt, we tested it against the 
best performing package known to the authors (id est SEx)
using a DPOSS field centered on the North Galactic Pole.
We want also to stress here that - in order to have an objective test and
at difference of what is currently done in literature - NExt was checked 
not against the performances of an arbitrarily choosen observer but rather 
against a much deeper catalogue of objects obtained from better quality 
material.

The comparison of NExt performances against those of SEx show that in the
detection phase, NExt is at least as effective as SEx in detecting ``true"
objects but much cleaner of spurious detections.  For what classification is
concerned, NExt NN performs better than the  SEX NN: 28 errors for NExt against 41
for SEx on a total of 460 objects, most of the errors referring to objects fainter
than the plate detection limit.

Other attempts, besides those described in the previous sections, to use
NN for similar tasks have been discussed in the literature. 
%
%
Balzell \& Peng (1998), used the same North Galactic Pole field  (but extracted
from POSS-I plates) used in this work. They tested their star/galaxy
classification NN on objects which are both too few (60 galaxies and 27 stars)
and too bright (a random check of their objects shows that most of the galaxies
extend well over than 20 pixels) to be of real interest.  It needs also to be
stressed that, due to their preprocessing strategy, their NN's are forced to
perform cluster analysis on a huge multidimensional imput space with scarsely
populated samples.

Naim (1997) follows instead a strategy which is similar to ours and makes use of a
fairly large dataset  extracted from POSS-I material. He, however, trained the
networks to achieve the same performances of an  experienced human observer while,
as already mentioned, NExt is checked against a catalogue of ''True" objects. Even
though his target is the classification of objects fainter and larger than those
we are dealing with, he tested the algorithm in a much more crowded 
and difficult region of the sky near the Galactic plane.  

O92 makes use of a traditional MLP and succeeded in demonstrating that 
AI methods can reproduce the star/galaxy classification obtained with traditional diagnostic diagrams by 
trained astronomers. Their aim, however, was less ambitious than that of ``performing the correct star/galaxy 
classification" which is instead the final goal of NExt.

This paper is a first step toward the application of Artificial Intelligence 
methods to astronomy. Foreseen improvements of our approach are the use of ICA
(Independent Component  Analysis) NN's instead of PCA NN's and the adoption of
Bayesian learning techniques to improve the classification performences of MLP's.
These developments and the application of NExt to other wide field astronomical
data sets obtained at large format CCD detectors will be discussed in forthcoming
papers.

{\bf Acknowledgements} The authors wish to thank Chris Pritchet for providing them with a digital
version of the IP92 catalogue. We also acknoledge the Canadian Astronomy Data
Center for providing us with POSS-II material.
This work was partly sponsored by the special grant MURST COFIN 1998,
n.9802914427.


\newpage

\begin{table*}
  \caption{First transposed eigenvector $3 \times 3$}
  \begin{tabular}{@{}rrr@{}}
1      & 1     & 1  \\
1      & 1     & 1  \\
1      & 1     & 1
\end{tabular}
\end{table*}

\begin{table*}
  \caption{Second transposed eigenvector $3 \times 3$}
  \begin{tabular}{@{}rrr@{}}
-2      & -4     & -4  \\
1      & 0     & 1  \\
4      & 4     & 2
\end{tabular}
\end{table*}

\begin{table*}
  \caption{Third transposed eigenvector $3 \times 3$}
  \begin{tabular}{@{}rrr@{}}
4      & 0     & -3  \\
4      & 0     & -4  \\
3      & 0     & -4
\end{tabular}
\end{table*}

\begin{table*}
  \caption{First transposed eigenvector $5 \times 5$}
  \begin{tabular}{@{}rrrrr@{}}
0.184082       & 0.196004     & 0.192174   & 0.172190 &  0.135904 \\
0.207895        & 0.225043      & 0.223247    & 0.202225  &  0.161833  \\
0.216280  & 0.236501 & 0.236244 & 0.215310 &  0.173737 \\
0.207416 & 0.228742 & 0.229733 & 0.210264 &  0.170427 \\
0.181968 & 0.202628  & 0.204979  & 0.188628  &  0.153777
\end{tabular}
\end{table*}

\begin{table*}
  \caption{Second transposed eigenvector $5 \times 5$}
  \begin{tabular}{@{}rrrrr@{}}
0.333702 & 0.313831 & 0.211281 & 0.095904 & -0.015149 \\
0.340559 & 0.244218 & 0.181150 & 0.022373 & -0.018147 \\
0.230683 & 0.121065 & 0.006941 & -0.130654 & -0.200468 \\
0.052375 & -0.053280 & -0.130905 & -0.292818 & -0.308128 \\
-0.039431 & -0.078693 & -0.241601 & -0.257146 & -0.256896
\end{tabular}
\end{table*}

\begin{table*}
  \caption{Second transposed eigenvector $5 \times 5$}
  \begin{tabular}{@{}rrrrr@{}}
0.043911 & -0.140093 & -0.208548 & -0.245815 & -0.308660 \\
0.114738 & -0.053939 & -0.109611 & -0.273351 & -0.340794 \\
0.239802 & 0.210806 & 0.015948 & -0.166681 & -0.230214 \\
0.300927 & 0.256129 & 0.042729 & -0.009986 & -0.113216 \\
0.322900 & 0.244557 & 0.165326 & 0.004180 & -0.122482
\end{tabular}
\end{table*}

\newpage
\quad
\newpage
\quad
\newpage

\begin{table*}
  \caption{
Number of objects grouped in ``Total", ``True" and ``False" detections integrated
over the whole magnitude range. The reference catalog consists of 4819 objects, among which 
$\sim 2400$ are too faint to be visible on our plate material.}
  \begin{tabular}{@{}rrrr@{}}
Catalogues      & \multicolumn{3}{c}{Detections}\\
		&   total     & ``True'' objects & `False'' objects\\
K5              &   1942      & 1738     & 204  \\
MLNG3           &   2742      & 2059     & 683  \\
MLNG5           &   3776      & 2310     & 1466 \\
NG3             &   1584      & 1477     & 107  \\
NGCGS5          &   1862      & 1692     & 170 \\
SEx		&   4256       &   2388   & 1866 \\
\end{tabular}
\end{table*}

\begin{table*}
  \caption{Extracted features}
  \begin{tabular}{@{}rlrrlr@{}}
 {\it number} & {\it Features} & {\it Symbols} & {\it number} & {\it Features} & {\it Symbols} \\
 1      & Isophotal Area     & $A$  & 14 & Gradient $1-2$    & $G_{12}$    \\
 2      & Photometric Barycenter Abscissa & $\bar{x}$  & 15 & Gradient $2-3$          & $G_{23}$    \\
 3 & Photometric Barycenter Ordinate & $\bar{y}$ & 16 & Gradient $3-4$ & $G_{34}$  \\
 4 & Semimajor Axis & $a$  & 17 & Average Surface Brightness & $\langle SuBr \rangle$ \\
 5 & Semiminor Axis & $b$  & 18 & Central Intensity & $I_0$ \\
 6 & Position Angle & $\alpha$  & 19 & Ratio 1& $T_r$ \\
 7 & Object Diameter & $dia$  & 20 & Ratio 2& $T_{cA}$ \\
 8 & Ellipticity of the Object Boundary & $ell$ & 21 & Second Total Moment& $C_{2}$ \\
 9 & Filling Factor & $ f_{fac}$ & 22 & Fourth Total Moment& $C_{4}$ \\
10 & Area Logarithm & $ c_{2}$ & 23 & Ellipticity (Averaged over the whole & $E$ \\
11 & Armonic Radius & $ r_{-1}$ &  & Area)& \\
12 & Gradient $1-4$& $G_{14}$ & 24 & Peak Intensity & $I_p$ \\
13 & Gradient $1-3$& $G_{13}$ & 25 & Kron Radius & $r_{Kron}$
\end{tabular}
\end{table*}

\noindent
\begin{table*}
  \caption{Adopted sets of features.}
  \begin{tabular}{lrr}
  Source                          & code & number of the corresponding feature \\
                                &      & in Table 8 \\
\hline				
Odewhan full set                & O-F  & 7, 8, 17, 18, 9, 10, 14, 15, 16 \\
Odewhan large galaxies set      & O-L  & 17, 14, 15, 16 \\
Odewhan small galaxies set      & O-S  & 17, 18, 16, 7 \\
Godwin \& Peach like 1977 first & GP1  & 17, { \it flux}\\
Godwin \& Peach like 1977 second& GP2  & 10, { \it flux}\\
\end{tabular}
\end{table*}

\newpage
\quad
\newpage
\quad
\newpage

\begin{figure*}
\vskip 0.2 truecm
\caption{A portion of the {\it Hubble Deep Field}. The two panels show the
objects in the Couch (1996) and Lanzetta, Yahil \& Fernandez-Soto (1996) catalogs. The
figures are taken from Ferguson (1998). }
\end{figure*}

\begin{figure*}
\caption{Hierarchical PCA NN (left) and Symmetric PCA NN(right).}
\end{figure*}

\begin{figure*}
\caption{The structure of a hierarchical unsupervised NN.}
\end{figure*}

\begin{figure*}
\caption{The studied field. The field is $2000 \times 2000$ arcsec wide. North is up and East is
left.The image is binned at $4 \times 4$.}
\end{figure*}

\begin{figure*}
\psfig{figure=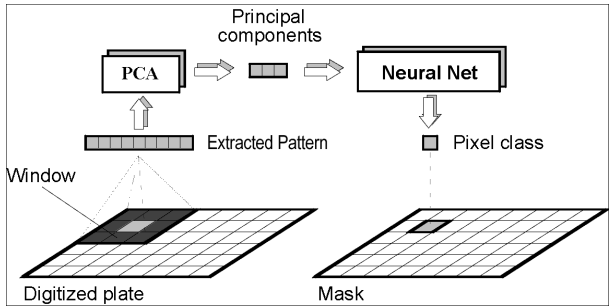,width=8truecm}
\caption{Single-band segmentation: process scheme.}
\end{figure*}

\begin{figure*}
\psfig{figure=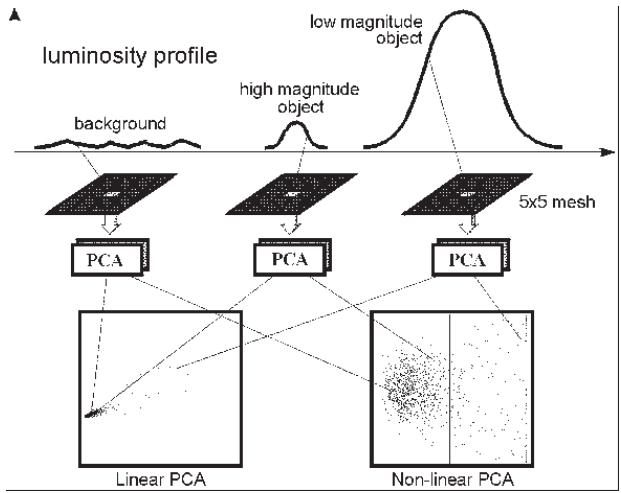,bbllx=0pt,bblly=0pt,bburx=578pt,bbury=474pt,width=8truecm}
\caption{Illustration of the coefficients distribution in 2 of 3 dimensional eigenvector space of
the input patterns. Nonlinear PCA NN's are the best performing because the spatial distribution 
corresponding imput are more spread out as shown in the example.}
\end{figure*}

\begin{figure*}
\caption{Clustering before and after the application of the $tanh$ function in
the learning of the PCA NN.}
\end{figure*}

\begin{figure*}
\centerline{\psfig{figure=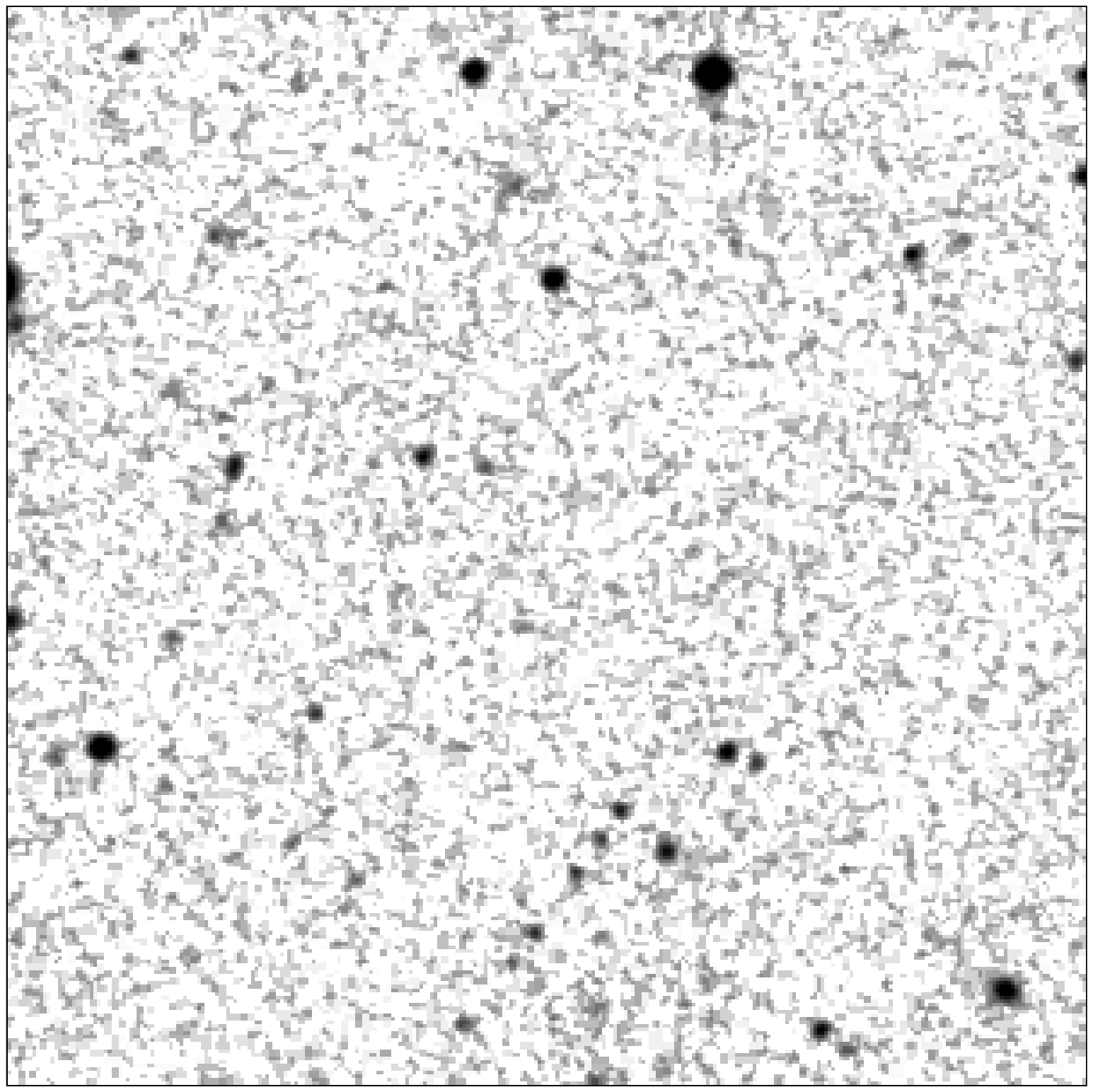,width=8cm} \psfig{figure=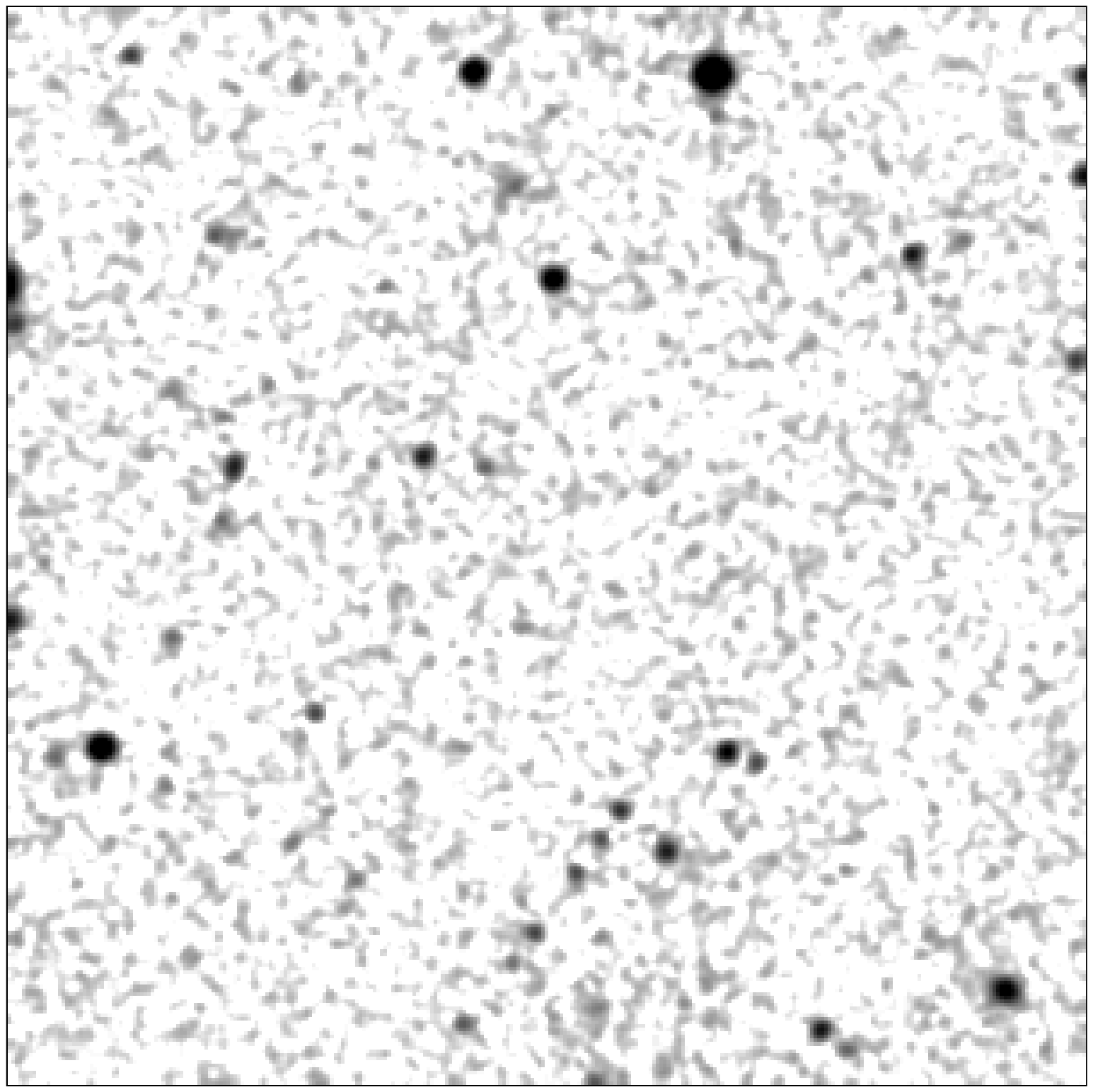,width=8cm}}
\centerline{\psfig{figure=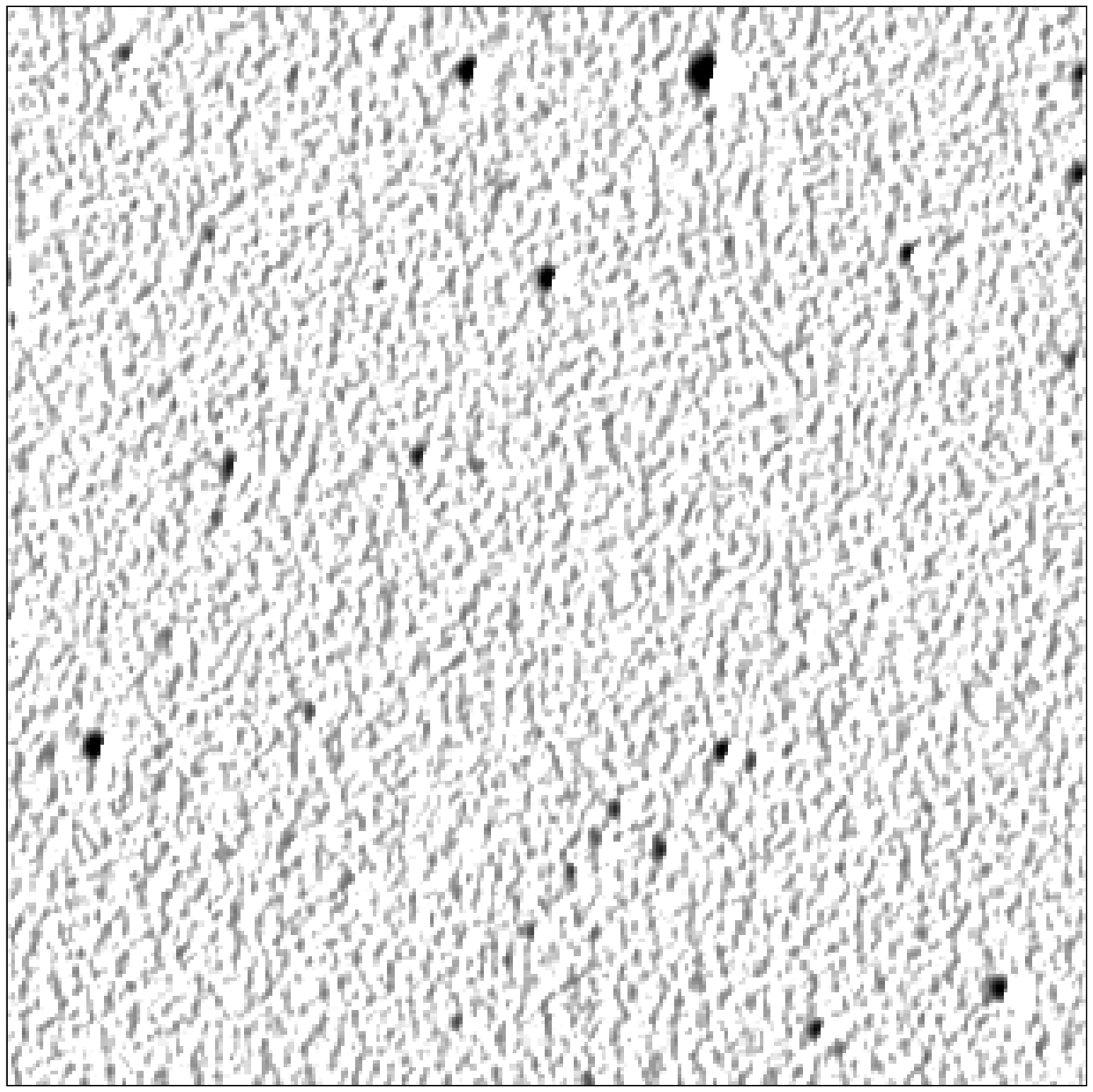,width=8cm} \psfig{figure=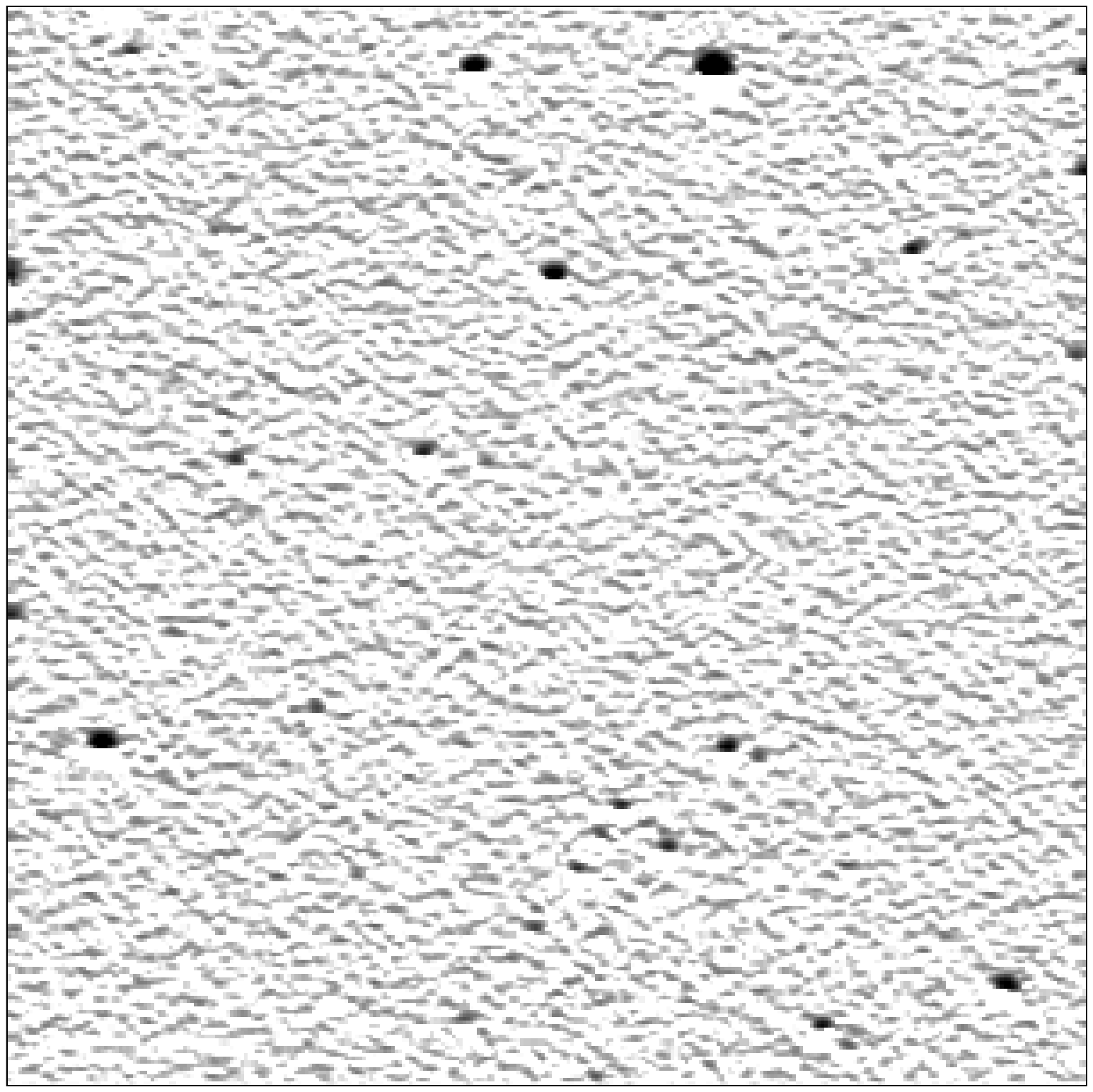,width=8cm}}
\caption{A portion $300 \times 300$ of the original image (top left). The same image convolved with
the principal eigenvector matrices: first eigenvector (top right), second and third eigenvectors
(bottom left and right, respectively).}
\end{figure*}

\newpage
\quad
\newpage
\quad
\newpage

\begin{figure*}
\psfig{figure=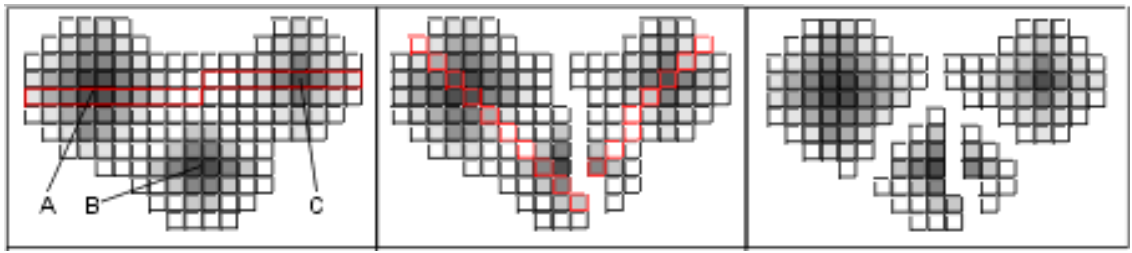,width=15truecm}
\caption{Deblending of three partially overlapping sources. In the left panel, 
the original image is
plotted. In the central and right panels, we show the result after the first and the second steps.
The correct result is achieved after a recomposition loop.}
\end{figure*}

\begin{figure*}
\psfig{figure=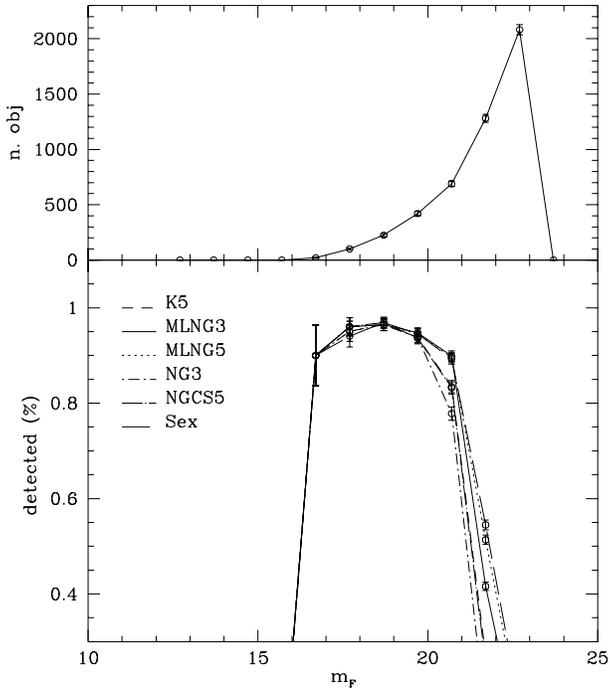,bbllx=55pt,bblly=195pt,bburx=470pt,bbury=685pt,width=8cm}
\caption{Upper panel: number of ``True" objects in the reference catalogue;
lower panel: number of objects detected by a given NN relative to the number of objects 
detected by SEx.}
\end{figure*}

\newpage
\quad
\newpage
\quad
\newpage

\begin{figure*}
\psfig{figure=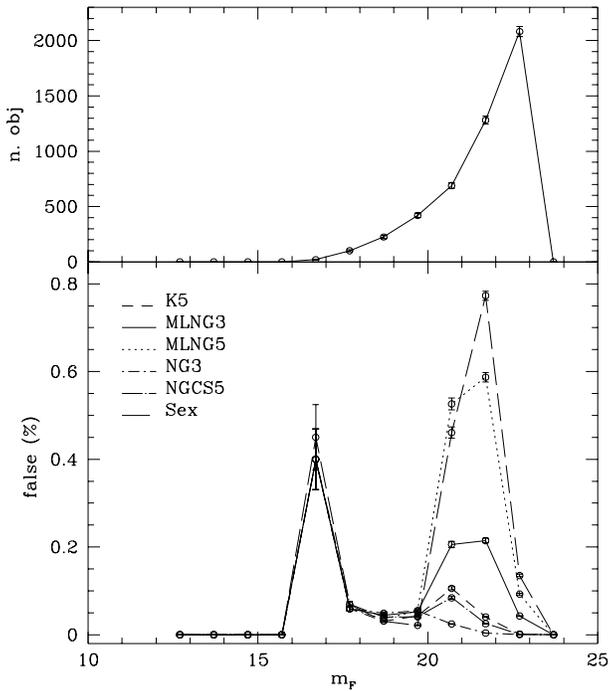,bbllx=55pt,bblly=195pt,bburx=470pt,bbury=685pt,width=8cm}
\caption{Upper panel: as in Fig. 10; lower panel fraction of "False'' objects detected 
by the algorithms but not present in the reference catalogue (IP92).}
\end{figure*}


\begin{figure*}
\psfig{figure=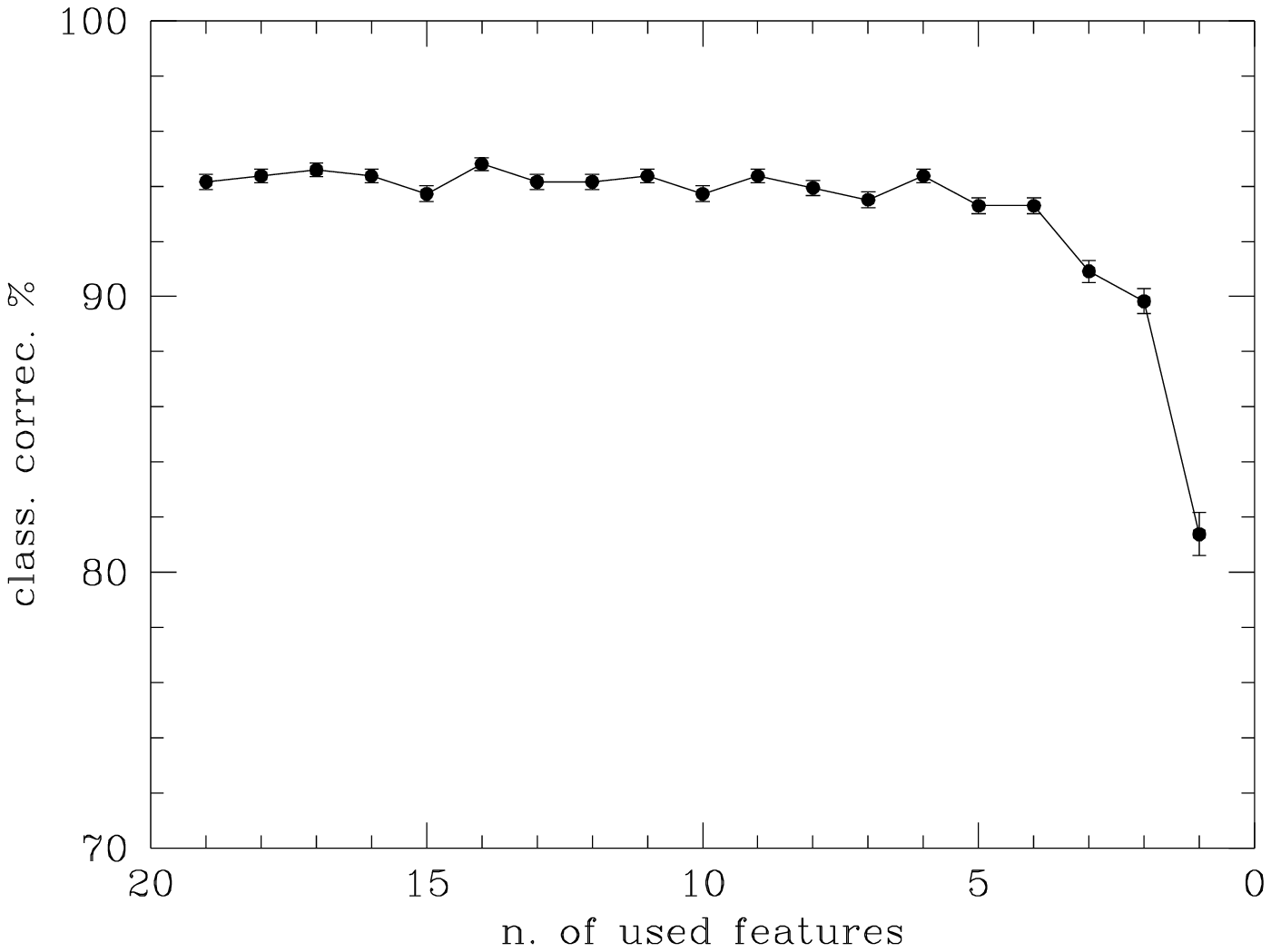,bbllx=65pt,bblly=195pt,bburx=465pt,bbury=630pt,width=8cm}
\caption{Percent number of objects detected by MLNG5, K5 and SEx.}
\end{figure*}

\begin{figure*}
\psfig{figure=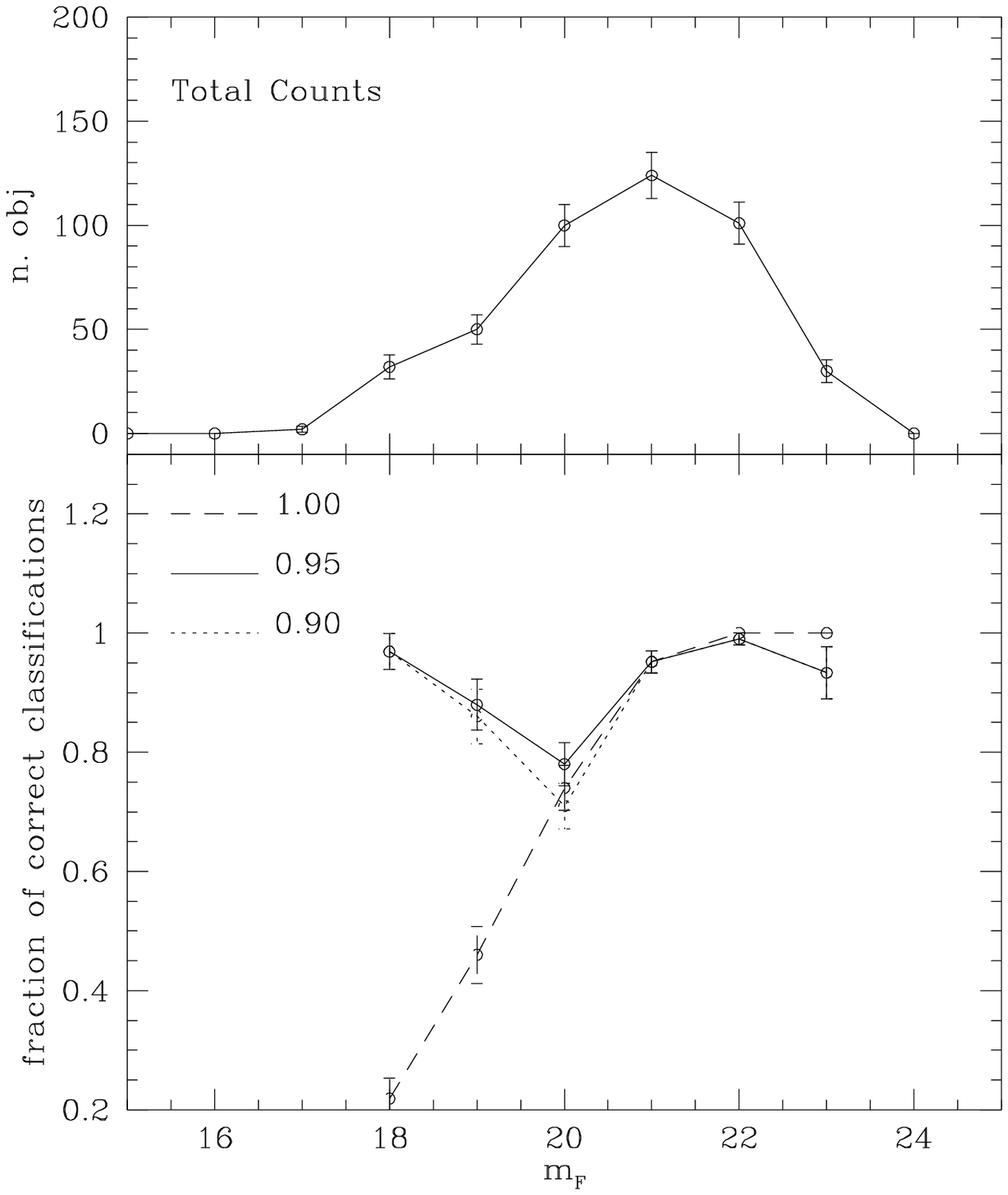,bbllx=55pt,bblly=195pt,bburx=465pt,bbury=505pt,width=8cm}
\caption{Classification performance as a function of the eliminated features.}
\end{figure*}

\begin{figure*}
\psfig{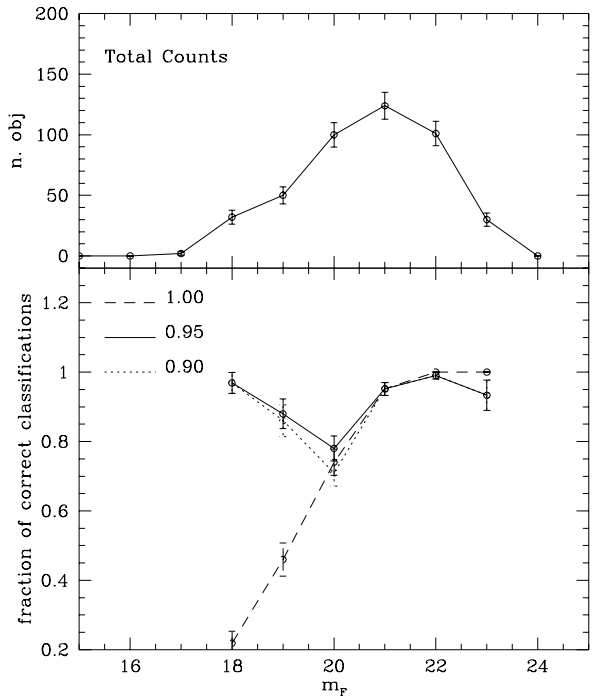}
\caption{Optimization of the classification performance of SEx for different
choises of the stellarity index parameter.
}
\end{figure*}

\newpage

\begin{figure*}
\psfig{figure=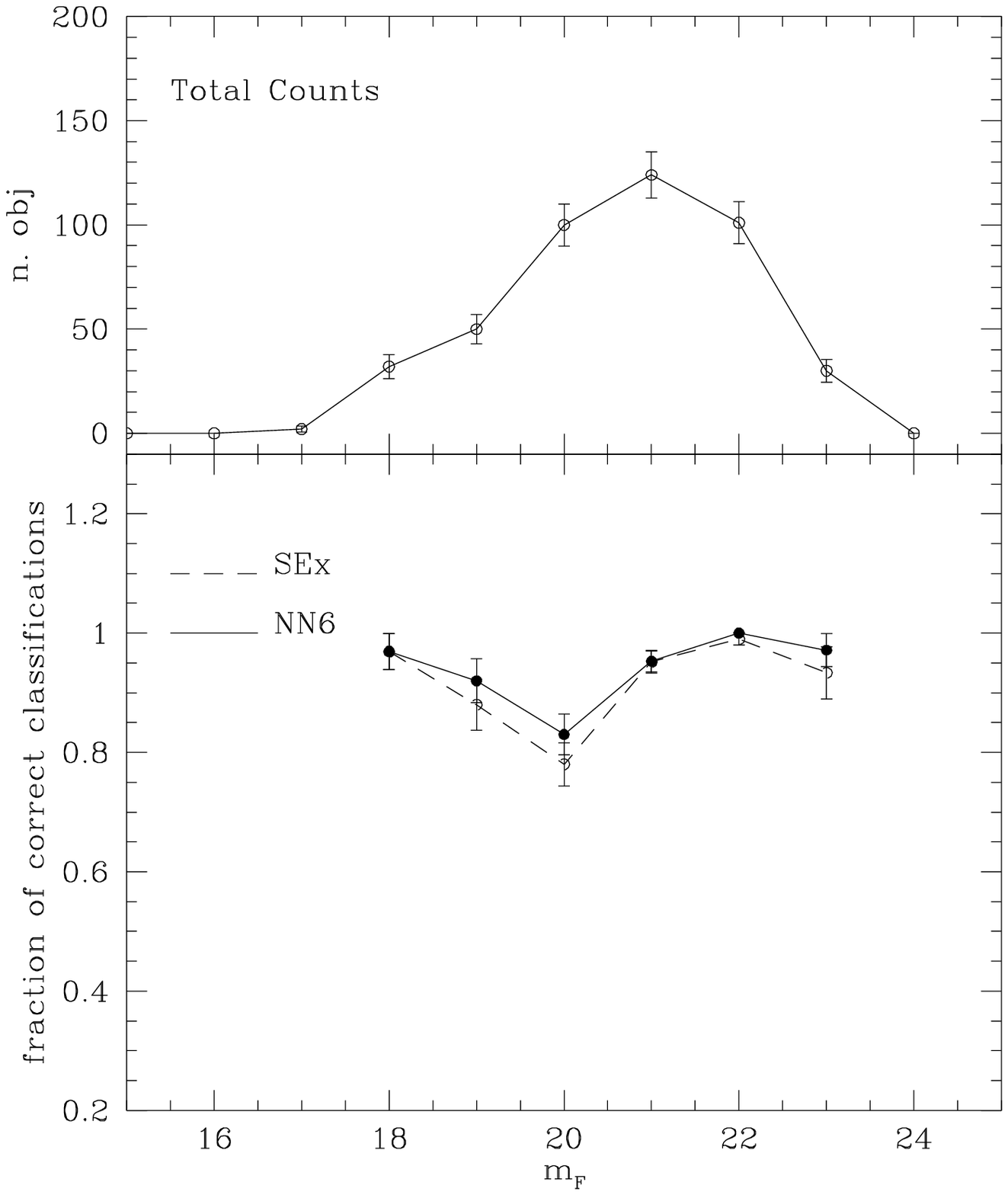,bbllx=55pt,bblly=195pt,bburx=470pt,bbury=685pt,width=8cm,width=9cm}
\caption{Classification performances of SEx and our NN based methods.}
\end{figure*}

\begin{figure*}
\psfig{figure=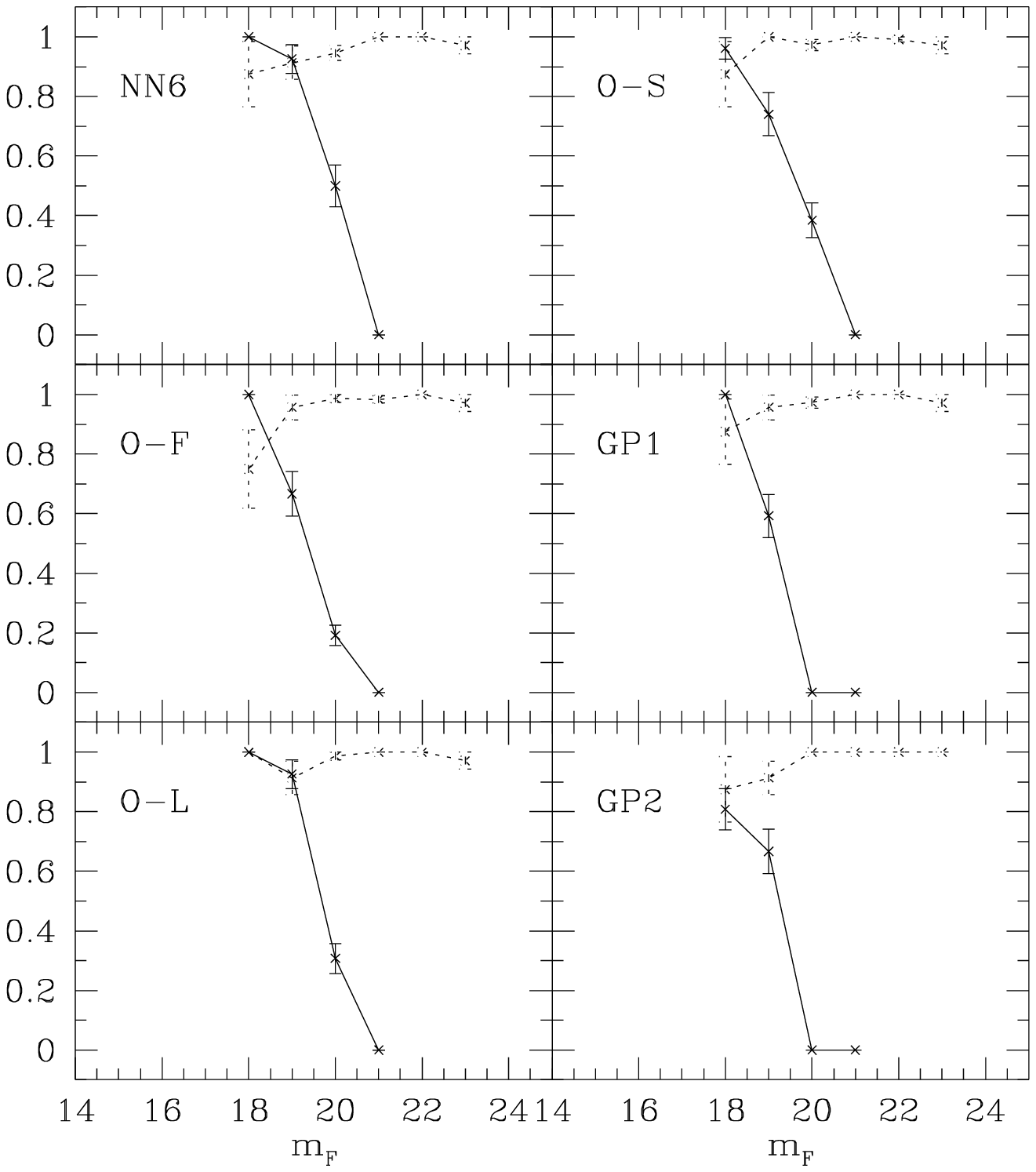,bbllx=75pt,bblly=240pt,bburx=465pt,bbury=675pt,width=8cm,width=9cm}
\caption{Classification performances of SEx and our NN based methods.}
\end{figure*}


\end{document}